\documentclass[%
 reprint,
superscriptaddress,
 amsmath,amssymb,
 aps,
]{revtex4-2}

\usepackage{graphicx}
\usepackage{dcolumn}
\usepackage{bm}

\begin{document}

\preprint{APS/123-QED}

\title{Optimal state for a Tavis-Cummings quantum battery via Bethe ansatz method}
\author{Wangjun Lu}
\email{wjlu1227@zju.edu.cn}
\affiliation{Zhejiang Institute of Modern Physics, Department of Physics, Zhejiang University, Hangzhou 310027, China }
\affiliation{Department of Maths and Physics, Hunan Institute of Engineering, Xiangtan 411104, China}

\author{Jie Chen}
\affiliation{Zhejiang Institute of Modern Physics, Department of Physics, Zhejiang University, Hangzhou 310027, China }

\author{Le-Man Kuang}
\email{lmkuang@hunnu.edu.cn}
\affiliation{Key Laboratory of Low-Dimensional Quantum Structures and Quantum Control of Ministry of Education,
Department of Physics and Synergetic Innovation Center for Quantum Effects and Applications,
Hunan Normal University, Changsha 410081, China }

\author{Xiaoguang Wang}
\email{xgwang1208@zju.edu.cn}
\affiliation{Zhejiang Institute of Modern Physics, Department of Physics, Zhejiang University, Hangzhou 310027, China }



\date{\today}

\begin{abstract}
In this paper, we investigate the effect of different optical field initial states on the performance of the Tavis-Cummings (T-C) quantum battery. In solving the dynamical evolution of the system, we found a fast way to solve the Bethe ansatz equation. We find that the stored energy and the average charging power of the T-C quantum battery are closely related to the probability distribution of the optical field initial state in the number states. We define a quantity called the number-state stored energy. With this prescribed quantity, we only need to know the probability distribution of the optical field initial state in the number states to obtain the stored energy and the average charging power of the T-C quantum battery at any time. We propose an equal probability and equal expected value splitting method by which we can obtain two inequalities, and the two inequalities can be reduced to Jensen's inequalities. By this method, we found the optimal initial state of the optical field. We found that the maximum stored energy and the maximum average charging power of the T-C quantum battery are proportional to the initial average photon number, and the quantum battery can be fully charged when the initial average photon number is large enough. We found two novel phenomena, which can be described by two empirical inequalities. These two novel phenomena imply the hypersensitivity of the stored energy of the T-C quantum battery to the number-state cavity field. Finally, we discussed the impact of decoherence on battery performance.
\end{abstract}

\maketitle


\section{introduction}
Large-capacity and high-density batteries have formed an inevitable requirement with the development of industrial technology, and high-density batteries inevitably require the miniaturization of battery cells. When the battery size is small enough that its dynamics are described by quantum mechanics, it is necessary to consider the impact of quantum effects on batteries. Based on this, a series of quantum batteries have been studied in recent years \cite{alicki2013entanglement,hovhannisyan2013entanglement,campaioli2017enhancing,manzano2018optimal,andolina2019quantum,zhang2019powerful,yang2020optimal,chen2020charging, PhysRevLett.122.047702,ferraro2018high,crescente2020ultrafast,le2018spin,peng2021lower,caravelli2020random,rossini2020quantum}, including Dicke quantum battery \cite{ferraro2018high,crescente2020ultrafast}, spin-chain quantum batteries \cite{le2018spin,peng2021lower}, Random quantum batteries \cite{caravelli2020random}, Sachdev-Ye-Kitaev Batteries \cite{rossini2020quantum}, topological batteries \cite{patil2020discharging}, and so on.  In considering the impact of quantum entanglement on quantum batteries, Alicki and Fannes found that global entangling operations could extract more work from a quantum battery than local operations \cite{alicki2013entanglement}. 
Hovhannisyan et al. found that a series of $N$ global entangling operations can extract the maximum work without creating any entanglement in the quantum battery \cite{hovhannisyan2013entanglement}. In studying the direction of the dissipative quantum battery,  F. Barra found that a unitary evolution with a single globally conserved quantity can extract work from the thermodynamic equilibrium state, and this process does not violate the second law of thermodynamics \cite{barra2019dissipative}.  J. Q. Quach and W. J. Munro use the dark states to charge and stabilize open quantum batteries. They found that the superextensive capacity and power density are correlated with entanglement, and the stored energy of the battery is stable without the need to continually access the battery \cite{quach2020using}. Since the decoherence of the quantum system will cause the aging of the quantum battery \cite{pirmoradian2019aging}, S. Y. Bai and J. H. An proposed to use Floquet engineering to suppress the decoherence of the quantum system to prevent the aging of the quantum battery \cite{bai2020floquet}.  

In Ref. \cite{PhysRevLett.122.047702}, the authors investigated the effect of three different cavity field initial states on the performance of the T-C quantum battery using the numerical or approximate method without considering other cavity field initial states. In this paper, we analytically solve the T-C model by the Bethe ansatz method to find the cavity field initial state that leads to the best performance of the T-C quantum battery. The Bethe ansatz (BA) is an exact method for the calculation of eigenenergies and eigenstates in quantum many-body systems, and it is a particular form of the wave function and was introduced by H. Bethe to solve the one-dimensional spin-1/2 Heisenberg model \cite{BetheH}. In the present, the method has been extended to other models in one dimension. Many other quantum many-body systems are known to be solvable by some variant of the Bethe ansatz \cite{lieb1963exact,lieb1963exact1,lieb1968absence,griffiths1964magnetization,CJBolech,NAndrei}. C. N. Yang used Bethe ansatz to study the Fermi problem and found that the prerequisite for the Fermi system to be solved is that it must satisfy a set of equations, which we now call Yang-Baxter equations \cite{yang1967some,baxter1972partition,takhtadzhan1979quantum}. The Yang–Baxter equation provides a set of relations that
can be solved to yield new integrable models. The BA can be constructed by using the Yang-Baxter algebra of the transfer matrix to generate the wavefunctions by applying quasi-particle creation operators to a pseudo-vacuum state. The BA is the quantum version of the Inverse Scattering Method \cite{korepin1997quantum}. Based on the Bethe ansatz method (BAM), N. M. Bogoliubov studies a series of integrable models in quantum optics \cite{bogoliubov1996exact,bogoliubov2013exactly,bogoliubov2017time}.
After constructing the BA of the system, the urgent problem is to solve the Bethe ansatz equation, a system of multivariate higher-order equations. Generally speaking, it is impossible to directly solve the Bethe ansatz equation for a large particle number system on a desktop computer. Now there are some software and methods for solving this equation, such as Bertini software \cite{BHSW06}, BertiniLab \cite{bates2016bertinilab}, Homotopy continuation method \cite{hao2013completeness,morgan2009solving}, and so on. However, using the above methods or software on a personal desktop computer cannot solve the Bethe ansatz equation of the large particle number system. In this paper, we propose a way to quickly solve the Bethe ansatz equation on a personal desktop computer.

In this paper, we use $N$ identical two-level atoms as the stored energy device and the cavity field as the energy supply device to study which cavity field initial state can improve the stored energy capacity and the average charging power of the QB. Here, the interaction between the atomic ensemble and the optical field is described by the Tavis-Cummings (T-C) model \cite{tavis1968exact}, so this stored energy device is called the T-C quantum battery (TCQB). We use the BAM to obtain the eigenvalues and eigenstates of the T-C Hamiltonian and then study the relationship between the probability distribution of the cavity field initial state in the number states and the performance of the TCQB. We propose an equal probability and equal expected value splitting method, which can help us find the optimal initial state of the TCQB. At the same time, we discovered two novel phenomena. These two phenomena with rich physical meaning can be described by two empirical inequalities, and their physical meaning implies that they have a wide range of applications in experiments. Finally, due to the unavoidable interaction between the cavity QED system and its environment in the experiment, which has been studied in spin squeezing \cite{li2021collective} and superradiation \cite{gegg2018superradiant, PhysRevLett.118.123602}, we investigate the effects of the decay of the optical field and the collective dephasing of the atoms on the performance of the TCQB.

\section{model and solution}

\begin{figure}[t]
\centering
\includegraphics[width=8.5cm,height=6cm]{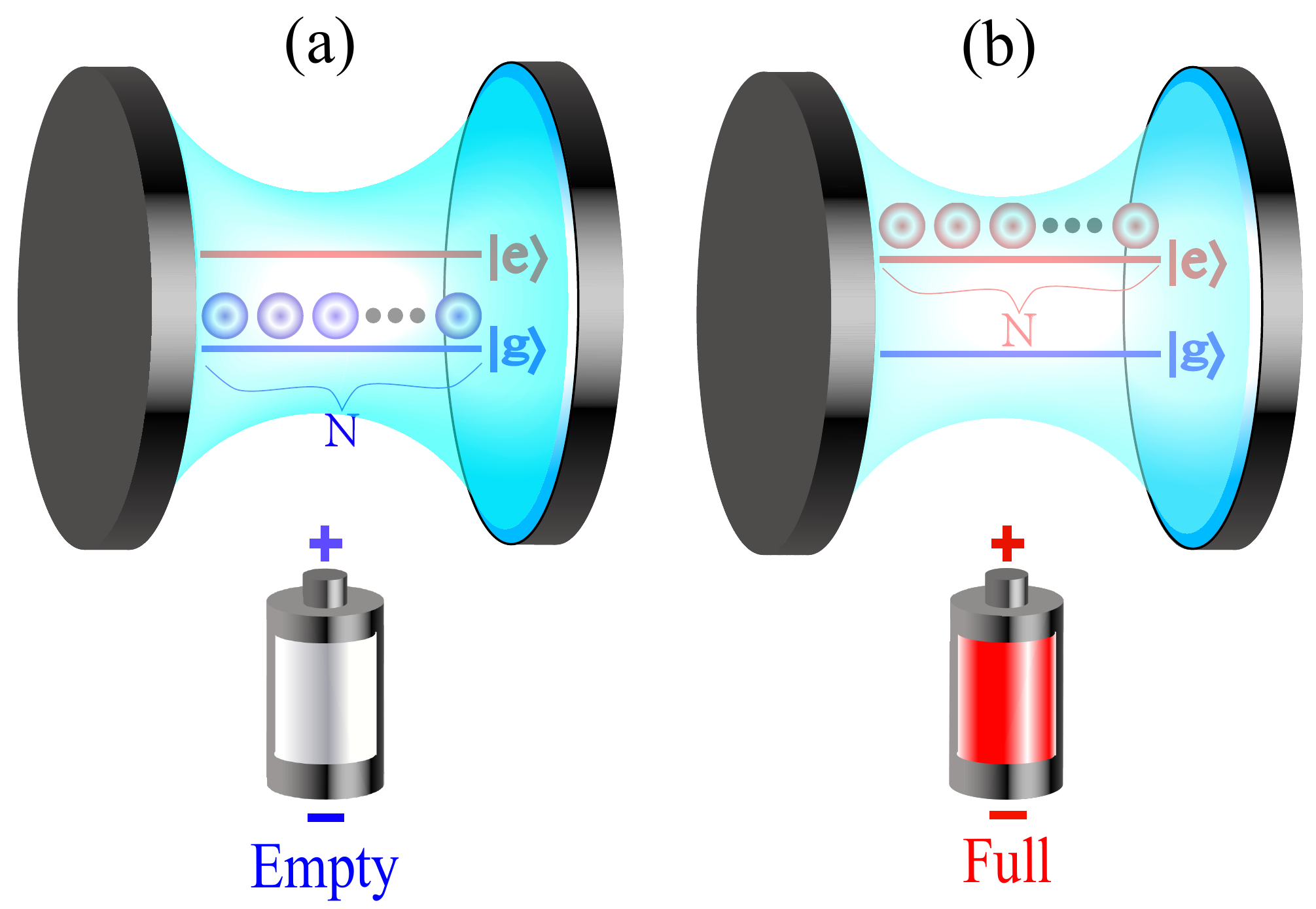}
\caption{\label{fig1}Schematic diagram of the T-C quantum battery. (a) The state of a quantum battery when it has no energy to output. All the atoms are in the ground state $\left|g\right\rangle$, and the battery is empty. (b) The quantum battery is in a fully charged state. $N$ identical atoms are all in the excited state $\left|e\right\rangle$.}
\end{figure}

As shown in Fig. \ref{fig1}, our system consists of a single-mode cavity field and $N$ identical two-level atoms. the cavity field and the atoms are coupled through electric dipole interaction, of which there is no interaction between the $N$ identical two-level atoms, and this model can be described by Dicke Hamiltonian ($\hbar =1$) \cite{dicke1954coherence,hepp1973superradiant}
\begin{equation}
\hat{H}_{D}=\omega_{c}\hat{a}^{\dagger}\hat{a}+\omega_{a}\hat{J}_{z}+
g(\hat{a}^{\dagger}+\hat{a})(\hat{J}_{+}+\hat{J}_{-}). \label{1}
\end{equation}
Here, $\omega_{c}$ and $\omega_{a}$ are the eigenfrequency of the cavity field and the transition frequency between the two energy levels of the atom, respectively. $\hat{a}^{\dagger}$ and $\hat{a}$ respectively represent  the creation and annihilation operators of bosons, which satisfy the commutation relations $[\hat{a},\hat{a}^{\dagger}]=1$, $[\hat{a},\hat{a}]=[\hat{a}^{\dagger}, \hat{a}^{\dagger}]=0$. $\hat{J}_{\alpha}=\Sigma_{i=1}^{N}\hat{\sigma}_{\alpha}^{i}/2$ in terms of the Pauli matrices $\hat{\sigma}_{\alpha}^{i} (\alpha=x,y,z)$ of the $i$-th atom is the collective angular momentum operator for the spin ensemble consisting of $N$ identical two-level atoms, these operators ${\hat{J}_{x}, \hat{J}_{y}, \hat{J}_{z}}$, satisfy the commutation relations of SU(2) algebra and $\hat{J}_{\pm}=\hat{J}_{x}\pm i\hat{J}_{y}$.

However, the Dicke model is not integrable, and its precise analytical solution cannot be obtained. The rotating-wave approximation (RWA) was introduced to overcome this problem. In the regime of near resonance $\omega_{c}\simeq\omega_{a}$ and relatively weak coupling $g\ll \min \{\omega_{c},\omega_{a}\}$, the  counter-rotating wave term can be ignored, and the Dicke model can be reduced to following T-C model
\begin{equation}
\hat{H}_{TC}=\omega_{c}\hat{a}^{\dagger}\hat{a}+\omega_{a}\hat{J}_{z}+
g(\hat{a}^{\dagger}\hat{J}_{-}+\hat{J}_{+}\hat{a}).  \label{2}
\end{equation}
The exact analytical solution of the T-C model exists. Under near resonance and weak coupling conditions, the Dicke model and the T-C model have the same dynamics in a short time. The primary purpose of our work is to find the optimal initial state of the cavity field so that the TCQB can achieve greater stored energy capacity and greater average charging power. Next, we will investigate the best initial state to achieve our goal by using the BAM.

In Appendix \ref{Appendix A}, we already know the eigenstates and eigenvalues of $\hat{H}_{TC}$, and now we study the charging problem of the TCQB in the eigen-representation of $\hat{H}_{TC}$. Let us start to consider the dynamic evolution of the system, the initial state of the system assume is 
\begin{equation}
\left|\Phi(0)\right\rangle =\left|\psi_{0}\right\rangle \otimes\left|J,-J\right\rangle  ,\label{3}
\end{equation}
where $\left|J,-J\right\rangle$ is the initial state of the atom ensemble, which means that all atoms are in the ground state. $\left|\psi_{0}\right\rangle$ is an arbitrary initial state of the cavity field. Then the state of the system at time $t$ is
\begin{equation}
\left|\Phi(t)\right\rangle =e^{-i\hat{H}_{TC}t}\left|\Phi(0)\right\rangle.\label{4} 
\end{equation}
Inserting the following unit operator in front of the initial state in Eq. (\ref{4})
\begin{equation}
\hat{I}_{M}=\sum_{\sigma}^{K}\frac{\left|\Phi_{J,M}\left(\{\lambda^{\sigma}\}\right)\right\rangle \left\langle \Phi_{J,M}\left(\{\lambda^{\sigma}\}\right)\right|}{N_{\sigma}^{2}},  \label{5}
\end{equation}
where $N_{\sigma}$ is the normalization coefficient of the $\sigma$-th eigenstate $\left|\Phi_{J,M}\left(\{\lambda^{\sigma}\}\right)\right\rangle $. Then, for a particular $M$, we can obtain the state of the system at time $t$ 
\begin{equation}
\left|\Phi_{M}(t)\right\rangle=\sum_{\sigma}^{K} \frac{C\left(\Phi_{J,M}^{\sigma},\Phi(0)\right)}{N_{\sigma}^{2}}  e^{-iE_{J,M}^{\sigma}t}\left|\Phi_{J,M}\left(\{\lambda^{\sigma}\}\right)\right\rangle ,  \label{6}
\end{equation}
where $C\left(\Phi_{J,M}^{\sigma},\Phi(0)\right)=\left\langle \Phi_{J,M}\left(\{\lambda^{\sigma}\}\right)\right|\left.\Phi(0)\right\rangle $, it is the inner product between the eigenstate of Hamiltonian $\hat{H}_{TC}$ and the initial state of the system, and its specific expression is  
\begin{equation}
C\left(\Phi_{J,M}^{\sigma},\Phi(0)\right)=C(M,\psi_{0})\sqrt{M!} . \label{7}
\end{equation}
The calculation of the above equation is shown in Appendix \ref{Appendix B}. Here, $C(M,\psi_{0})=\left\langle M |\psi_{0}\right\rangle $, and $(\left\langle M \right|)^{\dagger}$ is the number state which is the eigenstate of the number operator $\hat{a}^{\dagger}\hat{a}$.

\section{The stored energy and the average charging power of the TCQB}
The energy acquired by the atomic ensemble from the cavity field at time $t$ is
\begin{equation}
E(t)=\left\langle \Phi(t)\right|\hat{J}_{z}\left|\Phi(t)\right\rangle -
\left\langle \Phi(0)\right|\hat{J}_{z}\left|\Phi(0)\right\rangle  , \label{8}
\end{equation}
and the average charging power
\begin{equation}
P(t)=\frac{E(t)}{t}. \label{9}
\end{equation}
Since $\hat{M}=\hat{J}_{z}+\hat{a}^{\dagger}\hat{a}$, we can rewrite the above equation as follows
\begin{eqnarray}
E(t)&=&\left\langle \Phi(0)\right|\hat{M}\left|\Phi(0)\right\rangle -\left\langle \Phi(t)\right|\hat{a}^{\dagger}\hat{a}\left|\Phi(t)\right\rangle \nonumber \\
&&- \left\langle \Phi(0)\right|\hat{J}_{z}\left|\Phi(0)\right\rangle . \label{10}
\end{eqnarray}
Here we use $\left\langle \Phi(0)\right|\hat{M}\left|\Phi(0)\right\rangle$ instead of $\left\langle \Phi(t)\right|\hat{M}\left|\Phi(t)\right\rangle$ for $\hat{M}$ is a conserved quantity. Substituting the initial state $\left|\Phi(0)\right\rangle$ into the above equation, then 
\begin{equation}
E(t)=\bar{n}_{0} -\left\langle \Phi(t)\right|\hat{a}^{\dagger}\hat{a}\left|\Phi(t)\right\rangle, \label{11}
\end{equation}
where $\bar{n}_{0}=\left\langle \Phi(0)\right|\hat{a}^{\dagger}\hat{a}\left|\Phi(0)\right\rangle$ is the average photon number in the initial state. The physical meaning of the above equation is well understood. Since the total number of excitations is conserved, the total energy acquired by all atoms must be equal to the energy lost by the cavity field.
$\bar{n}_{0}$ is determined by the initial state of the cavity field, so we just need to obtain the average photon number in the cavity field at time $t$, and it is
\begin{equation}
\left\langle \Phi(t)\right|\hat{a}^{\dagger}\hat{a}\left|\Phi(t)\right\rangle=\sum_{M=0}^{\infty}\left\langle\Phi_{M}(t)\right|\hat{a}^{\dagger}\hat{a}\left|\Phi_{M}(t)\right\rangle. \label{12}
\end{equation}
Substituting Eq. (\ref{6}) into the above equation, then we obtain
\begin{equation}
\left\langle \Phi(t)\right|\hat{a}^{\dagger}\hat{a}\left|\Phi(t)\right\rangle=\sum_{M=0}^{\infty}\left|C(M,\psi_{0})\right|^{2}f(M,t), \label{13}
\end{equation}
where
\begin{eqnarray}
f(M,t)&=&M!\sum_{\gamma}^{K}\frac{1}{N_{\gamma}^{2}}e^{iE_{J,M}^{\gamma}t}\left\langle \Phi_{J,M}\left(\{\lambda^{\gamma}\}\right)\right|\hat{a}^{\dagger}\hat{a} \nonumber \\ &&\times\sum_{\sigma}^{K}\frac{1}{N_{\sigma}^{2}}e^{-iE_{J,M}^{\sigma}t}\left|\Phi_{J,M}\left(\{\lambda^{\sigma}\}\right)\right\rangle . \label{14}
\end{eqnarray}

In summary, the energy acquired by the atomic ensemble at time $t$ is
\begin{equation}
E(t)=\bar{n}_{0}-\sum_{M=0}^{\infty}\left|C(M,\psi_{0})\right|^{2}f(M,t). \label{15}
\end{equation}
The average photon number of the initial state can be expanded in the number state space 
\begin{equation}
\bar{n}_{0}=\sum_{M=0}^{\infty}\left|C(M,\psi_{0})\right|^{2}M, \label{16}
\end{equation}
then
\begin{equation}
E(t)=\sum_{M=0}^{\infty}\left|C(M,\psi_{0})\right|^{2}F(M,t), \label{17}
\end{equation}
and the average charging power
\begin{equation}
P(t)=\sum_{M=0}^{\infty}\left|C(M,\psi_{0})\right|^{2}\frac{F(M,t)}{t}, \label{18}
\end{equation}
here $F(M,t)=M-f(M,t)$, $\left|C(M,\psi_{0})\right|^{2}=\left|\left\langle M |\psi_{0}\right\rangle\right|^{2}$ is the probability distribution of the initial state $\left|\psi_{0}\right\rangle$  in the number states, and $F(M,t)$ is a binary function that varies with $M$ and $t$. From the above equation, we can see that the energy obtained by the atomic ensemble from the cavity field at time $t$ is actually an expected value of the function $F(M,t)$ under the probability distribution $\left|C(M,\psi_{0})\right|^{2}$. If we choose the initial state of the cavity field as the number state $\left|\psi_{0}\right\rangle=\left|M\right\rangle$, then
\begin{equation}
E(M,t)=F(M,t), P(M,t)=\frac{F(M,t)}{t}. \label{19}
\end{equation}
Here, we refer to $F(M,t)$ as the number-state stored energy. In fact, as long as we know the probability distribution of the optical field initial state in the number states, we can immediately know the stored energy and the average charging power of the TCQB at any time. In this way, we no longer need to solve the system dynamics when considering the influence of the other optical field initial state on the performance of the TCQB.

\begin{figure}[t]
\centering
\includegraphics[width=8.5cm,height=4.8cm]{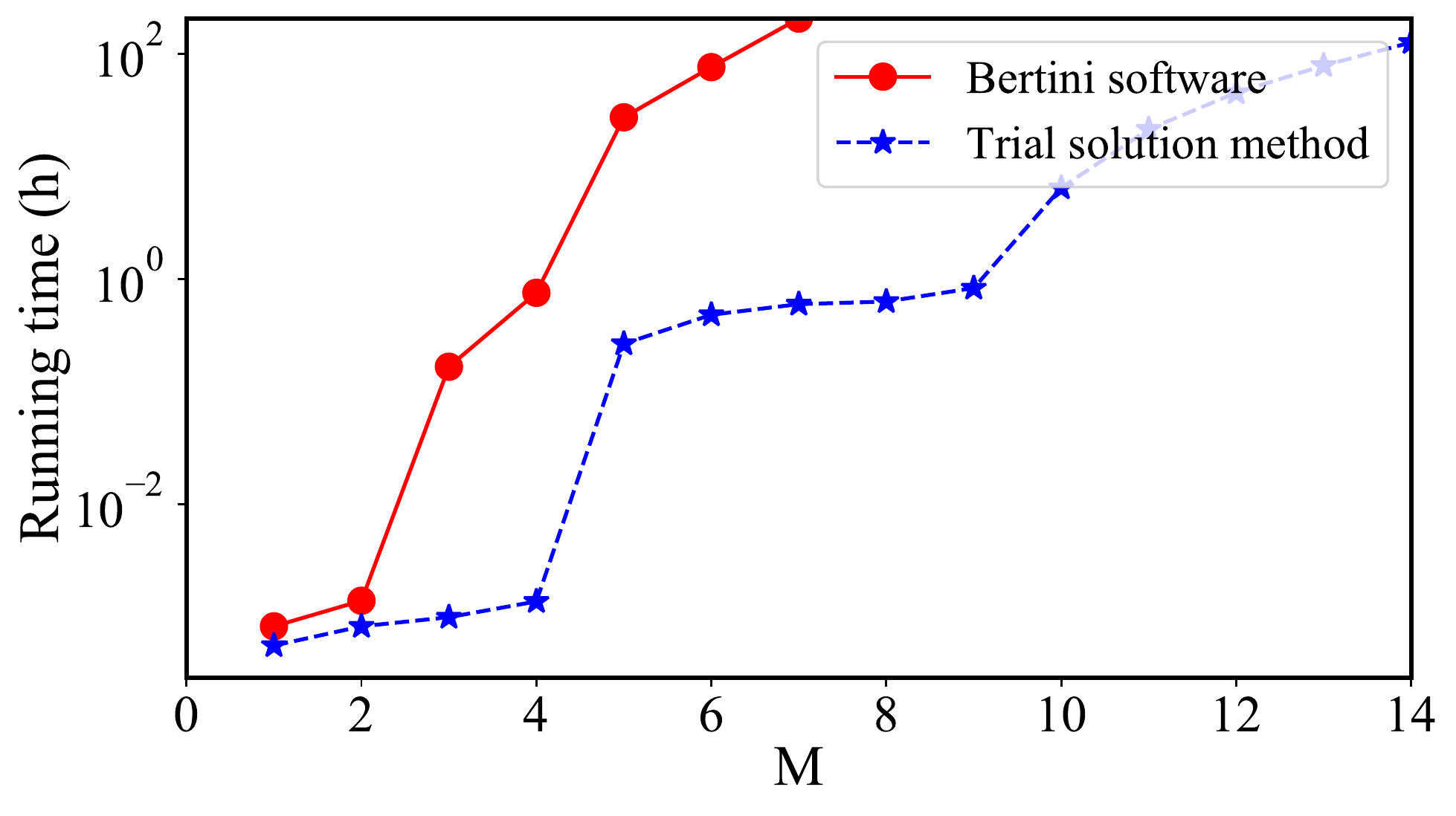}
\caption{\label{fig2}Running time as a function of the number $M$. Semilogarithmic plot giving the time (in hours) required to solve the BAE by the trial solution method (blue stars joined by
dashed lines) and by the Bertini software (red circles joined by solid lines). Here, we take $N=10$, and we did not use the Bertini software to solve the BAE when $M\geq8$ because it was too slow. }
\end{figure}

Next, to get the specific expression of $F(M,t)$, we only need to solve the BAE to obtain the solution of $\{\lambda^{\sigma}\}$. But generally speaking, it is challenging  to solve the BAE. Especially when $M$ is huge, we will face a high-order, multi-variable system of multiple equations, and the solution process is complicated. Fortunately, we have discovered a quick way to solve these equations called a trial solution method. The core idea is to use the value of the former set of solutions of BAE as the trial solution of the latter set of BAE. We consider a system with $10$ atoms to give a simple example. The Bethe ansatz equation is a two-variable quadratic equation system when $M=2$ and its three sets of solutions are easy to obtain. Here we assume that one set of the solution is $\lambda_{1}$, $\lambda_{2}$. If we need to solve the BAE when $M=3$, then we only need to take any one of $\{\lambda_{1}, \lambda_{2}\}$, $\lambda_{1}$, $\lambda_{2}$, and these three numbers $\{\lambda_{1}, \lambda_{1}, \lambda_{2}\}$ (or  $\{\lambda_{2}, \lambda_{1}, \lambda_{2}\}$) are used as the trial solution of the three-variable BAE when $M=3$, and then use Matlab to numerically solve these equations, We will quickly get the solutions. 

When $M$ becomes very large, the situation will start to become complicated. For example, when $M=8$, we will get $9$ sets of solutions, and we choose any one of them $\{\lambda_{1},\lambda_{2},\cdots,\lambda_{8}\}$, which contains $8$ values. When we need to solve the solution of BAE when $M=9$, we choose any one of $\{\lambda_{1},\lambda_{2},\cdots,\lambda_{8}\}$,  and then  form a set of trial solutions containing $9$ values with $\{\lambda_{1},\lambda_{2},\cdots,\lambda_{8}\}$ , we will get all the solutions. However, we may not find all the solutions. Namely, the number of solutions is not $10$. At this time, we need to randomly select $7$ values from $\{\lambda_{1},\lambda_{2},\cdots,\lambda_{8}\}$, and then randomly select $2$ values from $\{\lambda_{1},\lambda_{2},\cdots,\lambda_{8}\}$ to form a trial solution with $9$ values. If we find a new solution, but the number of solutions is still not $10$, then We respectively randomly select $6$ and $3$ values from $\{\lambda_{1},\lambda_{2},\cdots,\lambda_{8}\}$ to form a set of trial solutions with $9$ values, and so on. Finally, we will obtain all the solutions.

In Fig. \ref{fig2}, we compare the running time required to solve the Bethe ansatz equation for different values of $M$ by our trial solution method and by the Bertini software when $N=10$, and we find that the trial solution method has a significant advantage. When $M\geq8$, we did not use this software to solve the Bethe ansatz equation because the time required to solve the equation using the Bertini software was too long. Because these solutions of the BAE are too lengthy, in Appendix \ref{Appendix C}, we give some solutions of the BAE. We can get the corresponding binary function $F(M,t)$ through these solutions. Also, because these functions are cumbersome, we only list a few in Appendix \ref{Appendix C}.

\begin{figure}[t]
\centering
\includegraphics[width=8.5cm,height=14cm]{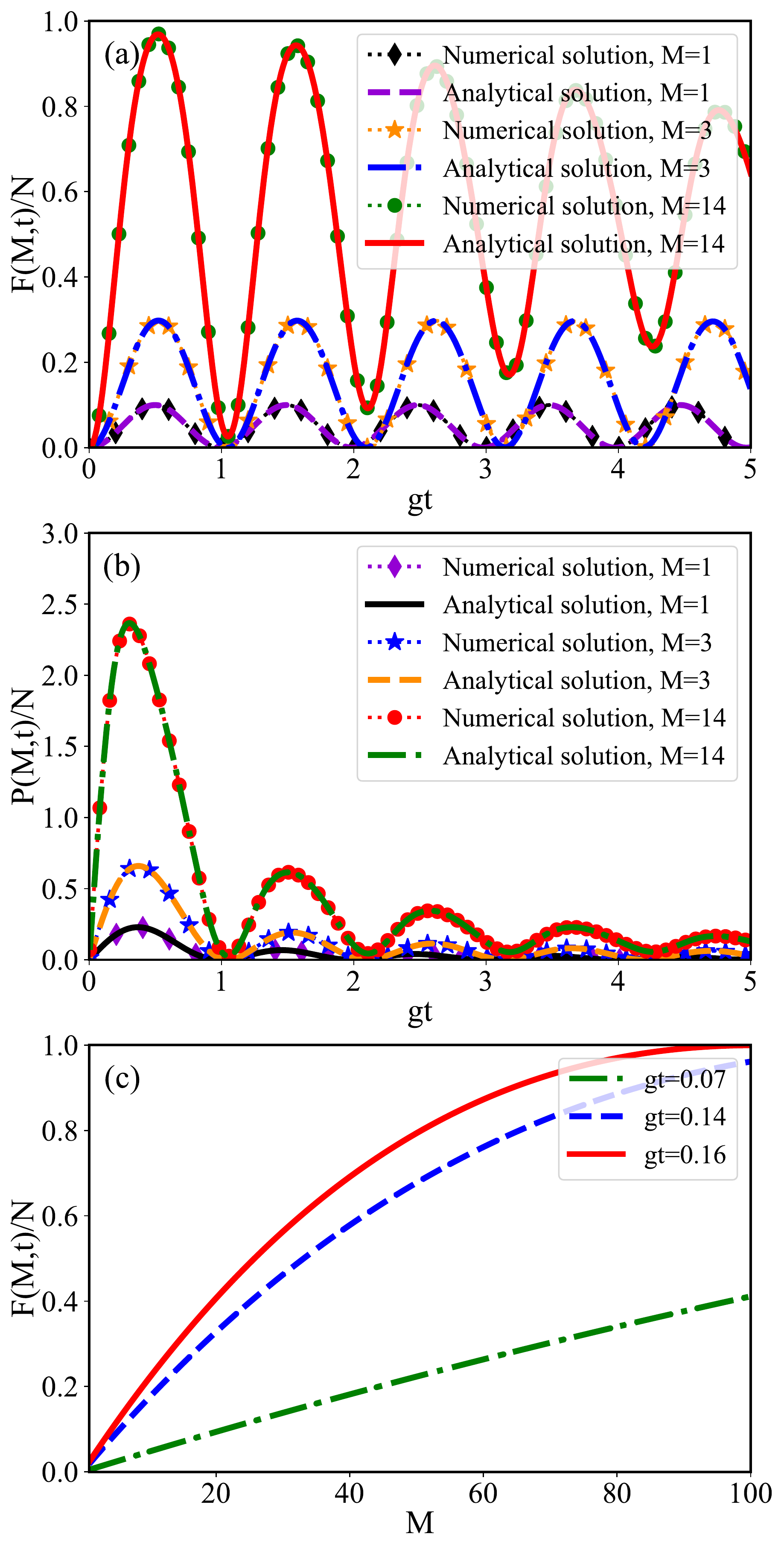}
\caption{\label{fig3} (a) The stored energy of the TCQB changes with time $t$ under different number states. (b) The average charging power of the TCQB varies with time $t$ under different number states. In (a) and (b), the marked lines and unmarked lines with different colors represent numerical and analytical solutions, respectively. (c) The energy acquired by the atomic ensemble varies with the average photon number $M$ of the number state at different times. In this paper, the number of atoms is $N=10$.}
\end{figure}

In Fig. \ref{fig3}(a), the energy obtained by the atomic ensemble from the cavity field changes with time for different $M$ are respectively drawn. The marked lines with different colors represent numerical solutions. The unmarked lines with different colors denote analytical solutions obtained with BAM. In Appendix \ref{Appendix C}, we give some concrete expressions of function $F(M,t)$, and we ignore some high-order small quantities in the analytical solution. The quantum toolbox QuTIP is used for numerical solutions \cite{johansson2012qutip}. It is worth noting that all marked lines are obtained by the numerical solution and all unmarked lines are obtained by the analytical solution should be strictly consistent in the power-related graphs. But we will find a little bit of deviation in some parts of the figure because we have already ignored some smaller values in $F(M,t)$. In Fig. \ref{fig3}(c), we have plotted the change of $F(M,t)$ with the initial average photon number $M$ at different times. From Fig. \ref{fig3}(a) and Fig. \ref{fig3}(c), we can intuitively see that $F(M,t)$ is proportional to $M$ in the time range we consider. Therefore, when the initial state of the cavity field is a number state, the number state with a larger initial average photon number can enable the TCQB to obtain more energy. Similarly, in Fig. \ref{3}(b), we can also see that the more the initial average photon number can improve the average charging power of the TCQB when the cavity field is in a number state in the time frame we consider. It is important to note that the time range we consider is from time $0$ to the time when the quantum battery first has the maximum stored energy in this article.

\begin{figure}[t]  
\centering
\includegraphics[width=8.5cm,height=14cm]{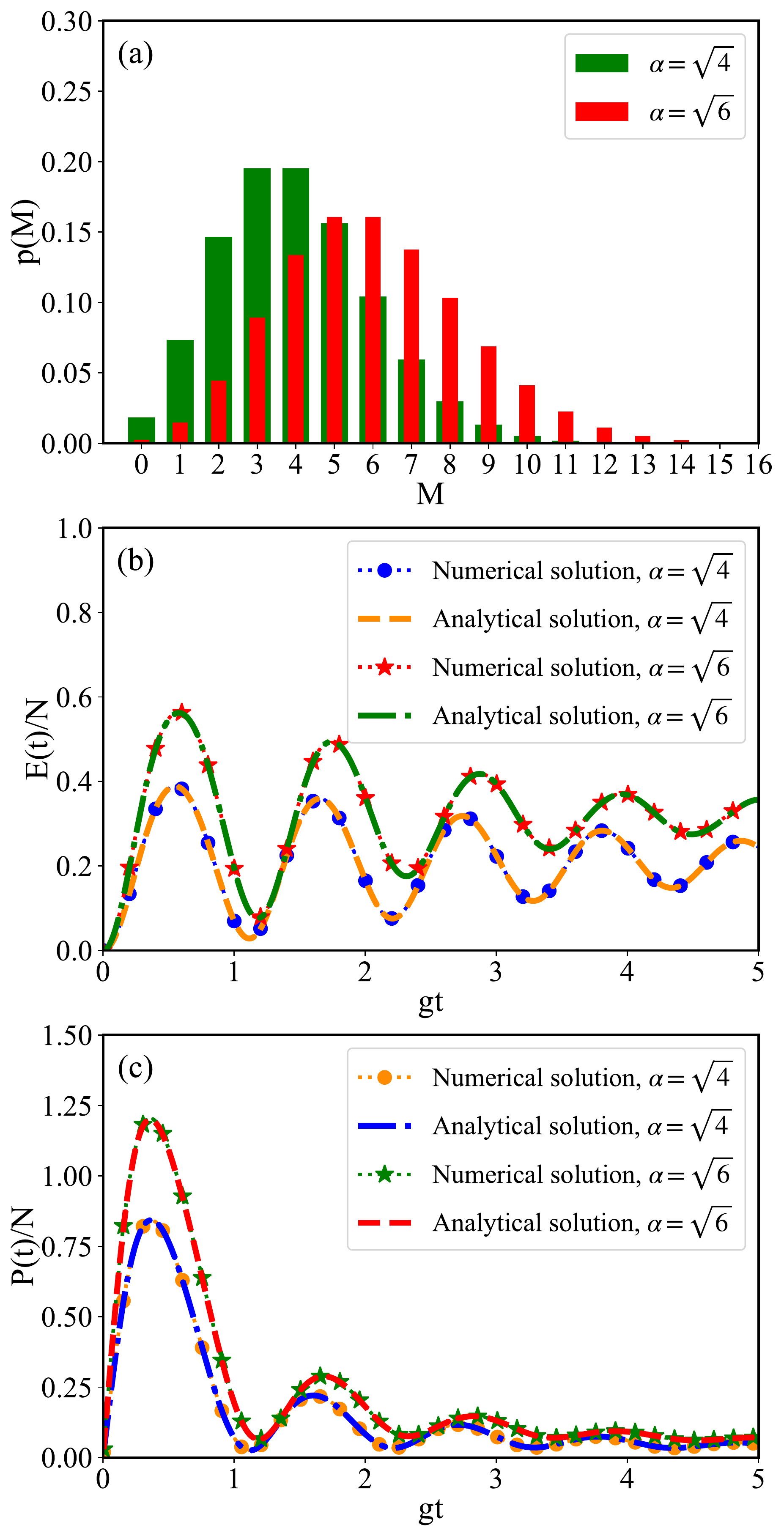}
\caption{\label{fig4}(a) The probability distributions of coherent states $\left|\sqrt{6}\right\rangle$ and $\left|\sqrt{4}\right\rangle$ in the number states, respectively. Here, we truncate the Hilbert space at the 16-photon Fock state $\left|M=16\right\rangle$. (b) The stored energy of the TCQB changes with time $t$ in the two initial coherent states, $\left|\sqrt{6}\right\rangle$ and $\left|\sqrt{4}\right\rangle$. (c) The average charging power of the TCQB varies with time $t$ in the two initial coherent states, $\left|\sqrt{6}\right\rangle$ and $\left|\sqrt{4}\right\rangle$. In (b) and (c), the marked lines and unmarked lines with different colors represent numerical and analytical solutions, respectively.}
\end{figure}

To verify whether the evolution of $E(t)$ and $P(t)$ with time $t$ in an arbitrary state of the initial cavity field is indeed the expected values of $F(M,t)$ and $F(M,t)/t$ under the probability distribution of the cavity filed initial state in the number states. We have studied the change of the stored energy and the average charging power of the TCQB with time when the initial state is the coherent state.
If the initial state is coherent state $\left|\alpha\right\rangle$, then the stored energy and the average charging power is
{\allowdisplaybreaks
\begin{eqnarray}
E(t)&=&\sum_{M=0}^{\infty}p(M)F(M,t) , \label{20}\\
P(t)&=&\sum_{M=0}^{\infty}p(M)\frac{F(M,t)}{t}, \label{21}
\end{eqnarray}}
where $p(M)=\left|\left\langle M |\alpha\right\rangle\right|^{2}$. In Fig. \ref{fig4}(a), we respectively draw the probability distribution of the coherent state in the number states when $\alpha=\sqrt{4}$ and $\alpha=\sqrt{6}$. According to these probability distributions and the function $F(M,t)$, the variations of the stored energy $E(t)$ and the average charging power $P(t)$ of the atomic ensemble with time $t$ are drawn respectively in Fig. \ref{fig4}(b) and Fig. \ref{fig4}(c). Indeed, the stored energy and the average charging power of the TCQB are the expected values of $F(M, t)$ and $F(M, t)/t$ under the probability distribution $p(M)$, respectively. In fact, not only for the coherent state but for any state, by finding its probability distribution in the number states and the function $F(M,t)$, we can draw the changes of the stored energy $E(t)$ and the average charging power $P(t)$ with time $t$.  From here, we can also see that the stored energy capacity and the average charging power of the TCQB are proportional to the initial average photon number. This is very easy to understand because, for the same type of state, the probability distribution of the state with more initial average photons is to the right in the number states. For example, when $\alpha=\sqrt{6}$, the probability distribution of the coherent state in the number states is more to the right than the probability distribution when $\alpha=\sqrt{4}$. We have proved that $F(M,t)$ increases as $M$ increases. Therefore, the larger the initial average photon number, the greater the stored energy capacity and the higher average charging power of the TCQB under the same type of the initial state.

\begin{figure}[t]
\centering
\includegraphics[width=8.5cm,height=14cm]{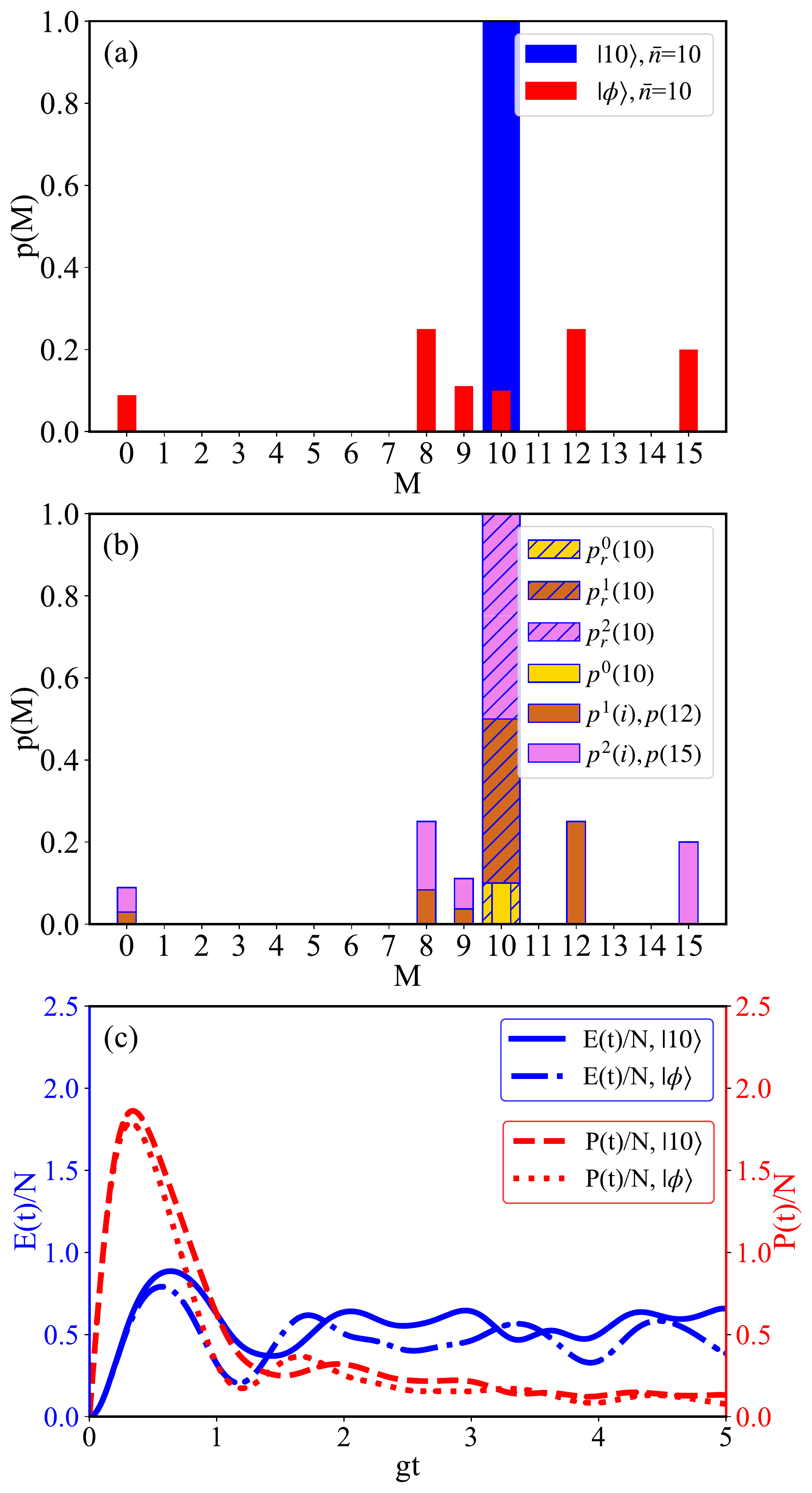}
\caption{\label{fig5} (a) The blue and red histograms respectively represent the probability distribution of number state $\left| 10 \right\rangle$ and the state $\left| \phi \right\rangle$ in the number states. (b) The probability distributions of the number state $\left| 10 \right\rangle$ and the state $\left| \phi \right\rangle$  after decomposition by the equal probability and equal expected value splitting method. (c) The variations of the stored energy and the average charging power of the TCQB with time $t$ under two different initial states, the number state $\left| 10 \right\rangle$ and the state $\left| \phi \right\rangle$.}
\end{figure}

\section{The Optimal initial state of the TCQB}

We found the relationship between the stored energy and the average charging power of the TCQB and the probability distribution of the optical field initial state in the number states. However, for an arbitrary cavity field initial state, it is uncertain whether the more the initial average photon number, the greater the stored energy capacity and the higher the charging power of the TCQB. In Appendix \ref{Appendix D}, under the constraint of equal average photon number, we found the optimal initial state using equal probability and equal expected value splitting method. To more easily understand our proposed equal probability and equal expected value splitting method, in the following, we use a concrete example to illustrate how to discover the optimal initial state.  

According to the equal probability and equal expected value splitting method, we assume that the initial average photon number $\bar{n}=10$. To prove that the number state is the optimal initial state for the maximum stored energy capacity and the maximum average charging power of the TCQB under the same initial average photon number, we use the number state $\left| 10 \right\rangle$ as the reference state and compare it with the following state 
\begin{eqnarray}
\left|\phi\right\rangle&=&\sqrt{p(0)}\left| 0 \right\rangle+\sqrt{p(8)}\left| 8 \right\rangle+\sqrt{p(9)}\left| 9 \right\rangle+\sqrt{p(10)}\left| 10 \right\rangle \nonumber \\
&&+\sqrt{p(12)}\left| 12 \right\rangle+\sqrt{p(15)}\left| 15 \right\rangle  .  \label{22}
\end{eqnarray}
We can give a set of probability distributions  $\{p(0)=4/45, p(8)=1/4, p(9)=1/9, p(10)=1/10, p(12)=1/4, p(15)=1/5 \}$ that satisfy the constraints. It is easy to verify that the number state $\left| 10 \right\rangle$ and the state $\left| \phi \right\rangle$ have the same average photon number. The probability distribution of the above two optical states in the number states is shown in Fig. \ref{fig5}(a). Our purpose is to compare the magnitude of the stored energy and the average charging power of the TCQB in these two initial states. The differences between them are as follows 
\begin{eqnarray}
\varDelta E(t)&=& F(10,t)- \Big[p(0)F(0,t)+\sum_{i=8}^{10}p(i)F(i,t) \nonumber \\ &&+p(12)F(12,t)+p(15)F(15,t)\Big]  ,\label{23}\\
\varDelta P(t)&=& \frac{\varDelta E}{t}  .  \label{24} 
\end{eqnarray}

\begin{figure}[t]
\centering
\includegraphics[width=8.5cm,height=5cm]{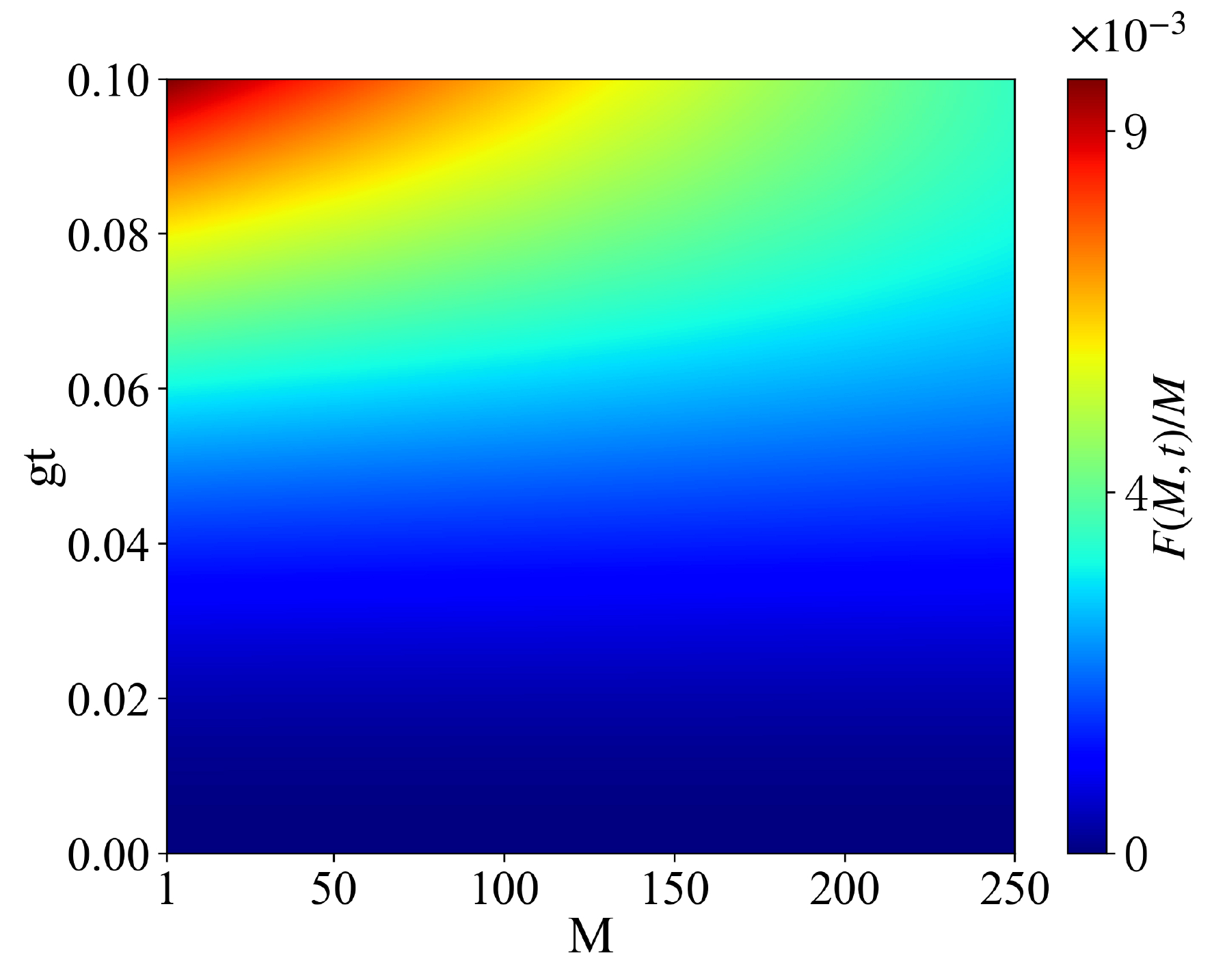}
\caption{\label{fig6}The average slope $F(M,t)/M$ as a function of the time $t$ and the average photon number $M$ of the initial number state.}
\end{figure}

Using the equal probability and equal expected value splitting method, we divide some probabilities into two parts $p^{1}(i)$ and $p^{2}(i)$ (where $i=2,8,9$), and divide the probability $p_{r}(10)$ of the number state $\left| 10 \right\rangle$ into three parts $p_{r}^{0}(10)$, $p_{r}^{1}(10)$, and $p_{r}^{2}(10)$. They satisfy the following relationships 
{\allowdisplaybreaks 
\begin{subequations}
\begin{eqnarray}
1&=&\sum_{j=0}^{2}p_{r}^{j}(10)  , \label{25a}\\
p(i)&=&\sum_{j=1}^{2}p^{j}(i), i=0,8,9 , \label{25b}\\
p_{r}^{0}(10)&=&p(10) ,\label{25c}\\
p_{r}^{1}(10)&=&\sum_{i=0,8,9}p^{1}(i)+p(12) , \label{25d}\\
p_{r}^{2}(10)&=&\sum_{i=0,8,9}p^{2}(i)+p(15) , \label{25e}\\
10p_{r}^{0}(10)&=&10p^{0}(10) ,\label{25f}\\
10p_{r}^{1}(10)&=&\sum_{i=0,8,9}ip^{1}(i)+12(12) , \label{25g}\\
10p_{r}^{2}(10)&=&\sum_{i=0,8,9}ip^{2}(i)+15p(15) . \label{25h}  
\end{eqnarray}
\end{subequations}}
In Appendix \ref{Appendix D}, we give a general solution of the probability of satisfying all the above equations, and the above separated probabilities are all shown in Fig. \ref{fig5}(b). Substituting  Eq. (\ref{25a})-Eq. (\ref{25h}) into Eq. (\ref{23}) gives
\begin{eqnarray}
\varDelta E(t)&=&\sum_{j=1}^{2}\Bigg[\sum_{i=0,8,9}p^{j}(i)\Big[F(10,t)-F(i,t)\Big]\Bigg] \nonumber\\
&&+\sum_{i=12,15}p(i)\Big[F(10,t)-F(i,t)\Big]. \label{26} 
\end{eqnarray}
Then we substitute the $p(12)$ and $p(15)$, which are obtained from Eq. (\ref{25d})-Eq. (\ref{25h}), into the above equation yields
\begin{eqnarray}
\varDelta E(t)&&=\sum_{i=0,8,9}(10-i)\Bigg[p^{1}(i)\Big[\frac{[F(10,t)-F(i,t)]}{10-i} \nonumber\\
&&-\frac{[F(12,t)
-F(10,t)]}{2}\Big]+p^{2}(i)\Big[\frac{[F(10,t)-F(i,t)]}{10-i}  \nonumber\\
&&-\frac{[F(15,t)-F(10,t)]}{5}\Big]\Bigg]. \label{27} 
\end{eqnarray}
Before going further, it is important to note that the time range we discuss is from time $0$ to the time when the TCQB reaches its maximum value for the first time. In Fig. \ref{fig3}(c) and Fig. \ref{fig6}, we can find that the average slope of $F(M,t)$ at any time decreases or remains constant as $M$ increases in the time range we consider. And since $10>i $, $p^{1}(i)>0$, and $p^{2}(i)>0$, $\varDelta E(t)\geq0$ and $\varDelta P(t)\geq0$, this indicates that the number state $\left| 10 \right\rangle$ enables the TCQB to have a larger stored energy capacity and a higher average charging power compared to $\left| \phi \right\rangle$. In Fig. \ref{fig5}{c}, we plot the variations of the stored energy and the average charging power of the TCQB with time $t$ for different cavity field initial states. From the actual dynamical evolution of the TCQB, we can obtain the same conclusion as above. In Appendix \ref{Appendix D}, we give complete proof that the optimal initial state that enables the TCQB to have the maximum stored energy capacity and the highest average charging power is the number state $\left| \bar{n} \right\rangle$ when the initial average photon number $\bar{n}$ is an integer. However, when the initial average photon number $\bar{n}$ is a non-integer, the optimal initial state is $\sqrt{1-(\bar{n}_{0}-[\bar{n}_{0}])}\left|[n_{0}]\right\rangle+\sqrt{\bar{n}_{0}-[\bar{n}_{0}]}\left|[\bar{n}_{0}]+1\right\rangle$, which is a superposition of the number states $\left|[n_{0}]\right\rangle$ and $\left|[n_{0}]+1\right\rangle$, where $[\bar{n}]$ indicates the integral part of $\bar{n}$.

\begin{figure}[t]
\centering
\includegraphics[width=8.5cm,height=9.3cm]{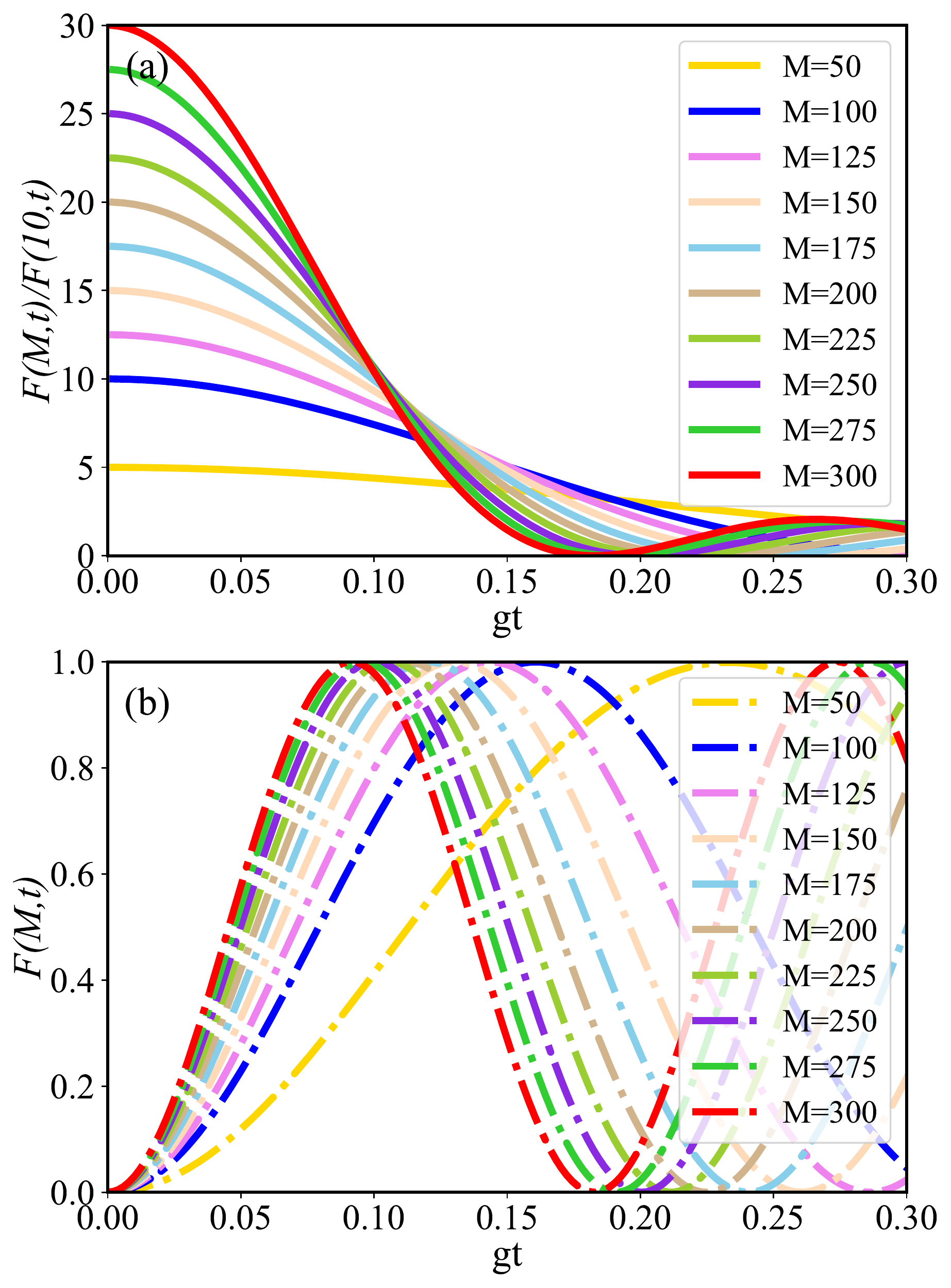}
\caption{\label{fig7}  The variations of $F(M,t)/F(10,t)$ and $F(M,t)$ with time $t$ when $M$ take different values are shown in (a) and (b), respectively.}
\end{figure}

\section{The hypersensitivity of the TCQB to the number-state optical field}

When we study the ratio of the energy obtained by the TCQB from the different number-state optical fields as a function of time $t$, we find two novel phenomena, which are summarized as follows using two empirical inequalities
\begin{eqnarray}
\frac{F(M,t)}{F(m,t)}&\leq&\frac{M}{m} ,\label{28} \\
\frac{d}{dt}\left[\frac{F(M,t)}{F(m_{0},t)}\right]&\leq&\frac{d}{dt}\left[\frac{F(m,t)}{F(m_{0},t)}\right] . \label{29}
\end{eqnarray}
The above two inequalities hold when $M\geq m\geq m_{0}$ and $t\leq t_{max}$ ($t_{max}$ represents the time when $F(M,t)$ takes the first maximum value, and show in Fig. \ref{fig7}(b). The physical meaning of the above two inequalities is very clear. Inequality (\ref{28}) shows that the stored energy of the TCQB is very sensitive to the number-state cavity field. When there is no coupling between the atomic ensemble and the number-state cavity field, we cannot obtain information about the cavity field through the stored energy of the TCQB. However, when there is a coupling between the atomic ensemble and the number-state cavity field, we can obtain the information of the initial average photon number of the cavity field by measuring the stored energy of the TCQB. Moreover, the shorter the coupling time between the atomic ensemble and the number-state cavity field, the more accurate we can obtain the information of the cavity field. As a simple example, the atomic ensemble with all atoms in the ground state is coupled with a known number-state cavity field $\left| m \right\rangle$, and then we can obtain the stored energy $E(m)$ of the atomic ensemble. Nextly, We use another atomic ensemble with the same initial properties to couple with an unknown number-state cavity field $\left| M \right\rangle$, and obtain the stored energy $E(M)$ of the atomic ensemble by measuring, then we can immediately get the average photon number of the unknown number-state cavity field, i.e., $M=mE(M)/E(m)$ (This equation holds when the coupling time is short, and the shorter the time, the more accurate the result). Inequality (\ref{29}) shows that the measurement is accurate even for long-term coupling when $m$ is not much different from $M$. Also, we can see from Fig. \ref{fig7}(b) that the TCQB can be fully charged when the initial average photon number is large enough.

\section{The effect of the decoherence on the performance of the TCQB}

\begin{figure}[t]
\centering
\includegraphics[width=8.5cm,height=9.3cm]{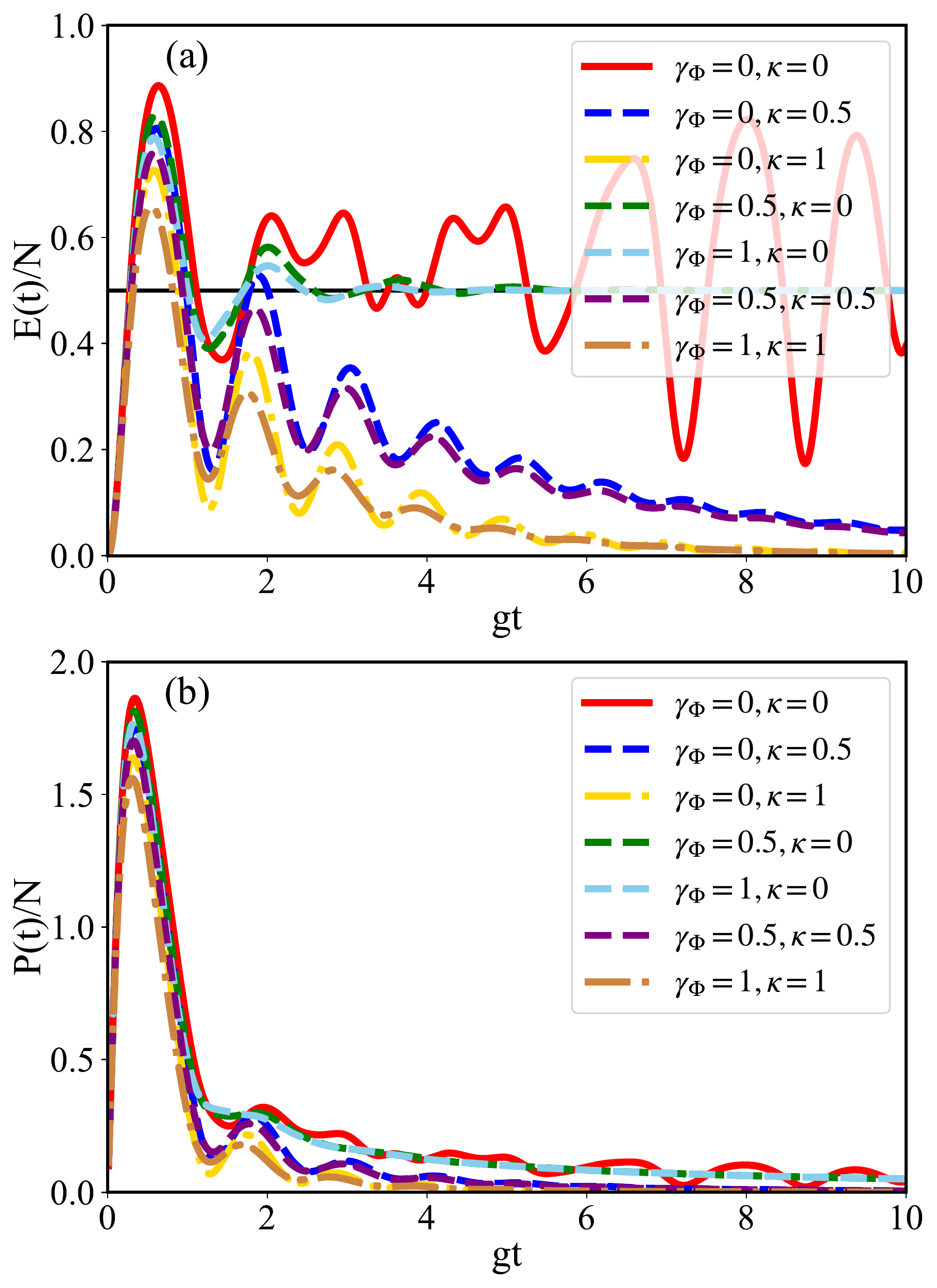}
\caption{\label{fig8} (a) The variations of the stored energy and the average charging power of the TCQB with time $t$ for the different collective dephasings of the atoms and the different decays of the optical field are shown respectively in (a) and (b). Here, the relevant parameters are chosen from Ref. \cite{PhysRevA.98.063815}, and all parameters are in unit of $\omega_{c}$. }
\end{figure}

Above, we have studied the optimal initial state of the TCQB in the process of unitary evolution. However, in practical applications, due to the interaction between the TCQB system and its environment, the impact of the decoherence on the performance of the TCQB must be considered. Based on this practical problem, we will study the effects of the collective dephasing of $N$ identical two-level atoms and the decay of the optical field on the performance of the TCQB when the TCQB is coupled with the environment. Under Markov approximation, Born approximation, and rotating wave approximation, the dynamics of the density matrix $\hat{\rho}$ of the TCQB is described by the following Lindblad master equation\cite{carmichael1999statistical}
\begin{eqnarray}
\frac{d\hat{\rho}}{dt}=-i[H_{TC}, \hat{\rho}]+\frac{\gamma_{\Phi}}{2}\mathcal{L}_{\hat{J}_{z}}[\hat{\rho}]+\frac{\kappa}{2}\mathcal{L}_{\hat{a}}[\hat{\rho}],   \label{30}
\end{eqnarray}
where the Lindblad superoperators are defined by $\mathcal{L}_{\hat{A}}[\hat{\rho}]=2\hat{A}\hat{\rho}\hat{A}^{\dagger}-\hat{A}^{\dagger}\hat{A}\hat{\rho}-\hat{\rho}\hat{A}^{\dagger}\hat{A}$, $\gamma_{\Phi}$ and $\kappa$ are the collective dephasing of the atoms and the decay of the cavity field, respectively. Generally, it is not easy to solve Eq. (\ref{30}) analytically, but the very friendly QuTIP \cite{johansson2012qutip} and PIQS \cite{PhysRevA.98.063815} packages  make it very convenient to solve the above density matrix. Then we can get the density matrix of the TCQB at any time. In Fig. \ref{fig8}, the time evolution of the stored energy of the TCQB, Fig. \ref{fig8}(a), and the average charging power of the TCQB, Fig. \ref{fig8}(b), are shown under the dynamics of Eq. (\ref{30}) for different system parameters.

In Fig. \ref{fig8}(a) and Fig. \ref{fig8}(b), at the initial time, $N$ identical two-level atoms are all in the ground state, and the light field is in the number state $\left|M=10\right\rangle$. We compared the effects of the different collective dephasings of the atoms and the different decay intensities of the optical field on the stored energy and the average charging power of the TCQB. We found that the collective dephasing of the atoms or the decay of the optical field can destroy the maximum stored energy and the maximum average charging power of the TCQB, and this destruction increases as the collective dephasing or decay increases. As photons continue to flow into the environment, the absorption and radiation of photons by the atoms no longer remain in balance. Finally, due to the complete leakage of the energy, the stored energy of the TCQB will also be exhausted. Especially under the limit of bad cavity ($\kappa\gg g$), the energy loss of the TCQB is faster, as shown by the yellow and brown dash-dot lines in Fig. \ref{fig8}(a). Since the decay of the cavity field destroys the performance of the TCQB, we need to avoid it as much as possible in practice. However, the difference is that although the collective dephasing of the atoms will destroy the maximum stored energy and the maximum average charging power of the TCQB, without the decay of the cavity field, the collective dephasing of the atoms will accelerate the evolution of the atoms to reach a steady state, which means the absorption and radiation of the photons by the atoms reach a dynamic balance, as shown by the green and sky-blue dashed lines in Fig. \ref{fig8}(a), and the solid black line represents the above-mentioned balance line.

\section{\label{Sec:4} Conclusion }
In this work, we mainly study the influence of the cavity field initial state on the stored energy and the average charging power of the T-C quantum battery. The purpose is to find the optimal initial state of the cavity field that enables the T-C quantum battery to have the maximum stored energy and the maximum average charging power. We use the Bethe ansatz method to find that the stored energy and the average charging power of the T-C quantum battery are closely related to the probability distribution of the cavity field initial state in the number states. We have defined a quantity called number-state stored energy. Through this quantity, we only need to know the probability distribution of the cavity field initial state in the number states to obtain the stored energy and the average charging power of the T-C quantum battery at any time. When solving the Bethe ansatz equation, we found a quick way to solve this equation. When studying the influence of the initial average photon number on the performance of the T-C quantum battery, we found that the larger the average photon number, the greater stored energy capacity and the higher average charging power of the T-C quantum battery. When studying the effect of the initial state of the different cavity fields on the T-C quantum battery under the same average photon number, we proposed an equal probability and equal expected value splitting method. In this way, we prove two inequalities that can be reduced to Jensen's inequalities. By these two inequalities we proved, and the average slope of the number state stored energy of the T-C quantum battery decreases as the initial photon number increases, we found the optimal initial state of the T-C quantum battery. Meanwhile, We found two novel phenomena, which imply the super-sensitivity of the stored energy of the T-C quantum battery to the number-state cavity field, and we have pointed out the potential applications of these two phenomena. Finally, we considered the effect of the decoherence on the performance of the T-C quantum battery. We find that both the collective dephasing of the atoms and the decay of the cavity field can reduce the maximum stored energy of the T-C quantum battery. However, the collective dephasing of the atoms can accelerate the evolution of the atoms to reach the steady state in the absence of the decay of the cavity field.

\begin{acknowledgments}
X. G. W was supported by the NSFC through Grant No.~11935012 and No.~11875231, the National Key Research and Development Program of China (Grants No.~2017YFA0304202 and No.~2017YFA0205700). W. J. L. was supported by the NSFC ( Grants No. 11947069) and the Scientific Research Fund of Hunan Provincial Education Department ( Grants No. 20C0495). L.-M.K. was supported by the NSFC
(Grants No. 11935006 and No. 11775075). 
\end{acknowledgments}

\appendix

\section{\label{Appendix A}A brief review of the eigenstates and eigenenergies of the T-C model }
To make our model and calculation clear, we first briefly review the exact analytical solution of the T-C model by BAM \cite{bogoliubov2017time}. Introducing three operators
\begin{eqnarray}
&&\hat{X}_{+}(\lambda)=\hat{a}^{\dagger}-\frac{\hat{J}_{+}}{\lambda},\hat{X}_{-}(\lambda)=\hat{a}-\frac{\hat{J}_{-}}{\lambda}, \nonumber  \\
&&\hat{X}_{z}(\lambda)=-\frac{\hat{J}_{z}}{\lambda}+\frac{\Delta-\lambda}{2},                  \label{A1}
\end{eqnarray}
and they satisfy some commutation relations shown in \cite{bogoliubov2017time}, $\lambda$ is a complex number and $\Delta=(\omega_{a}-\omega_{c})/g$. A new operator can be constructed by these three operators
\begin{equation}
\hat{t}(\lambda)=\left[\hat{X}_{z}(\lambda)\right]^{2}+\frac{1}{2}\hat{X}_{+}(\lambda)\hat{X}_{-}(\lambda)+\frac{1}{2}\hat{X}_{-}(\lambda)\hat{X}_{+}(\lambda).                \label{A2}
\end{equation}
Substituting Eq. (\ref{A1}) into Eq. (\ref{A2}), the specific expression of $\hat{t}(\lambda)$ can be obtained
\begin{equation}
\hat{t}(\lambda)=\lambda^{-2}\hat{J}^{2}-\lambda^{-1}\hat{H}_{N}+\hat{M}+\frac{(\Delta-\lambda)^{2}}{4}+\frac{1}{2} ,       \label{A3}
\end{equation}
where $\hat{J}^{2}$ is the square of total spin angular momentum, since it  commutes with the operators, which are any linear combination of $\{\hat{J}_{x}, \hat{J}_{y}, \hat{J}_{z}\}$, it commutes with $\hat{H}_{TC}$ $[\hat{J}^{2}, \hat{H}_{TC}]=0$. $\hat{M}=\hat{J}_{z}+\hat{a}^{\dagger}\hat{a}$ is the number of excitations.  $\hat{H}_{N}=(\hat{H}_{TC}-\omega_{c}\hat{M})/g=\Delta\hat{J}_{z}+\hat{a}^{\dagger}\hat{J}_{-}+\hat{J}_{+}\hat{a}$. It is easy to prove that the operators $\hat{M}$ and $\hat{H}_{TC}$ commute with each other $[\hat{M}, \hat{H}_{TC}]=0$, which implies the conservation of the total excitation number, then it is obvious that $\hat{H}_{N}$ and $\hat{H}_{TC}$ commute with each other as $[\hat{H}_{N}, \hat{H}_{TC}]=0$. In summary, $[\hat{t}_{\lambda}, \hat{H}_{TC}]=0$. Since two commutative operators with the same Hilbert space have the same eigenstates, if we find the eigenstates of $\hat{t}_{\lambda}$, then we can solve the dynamic problems of the system in the diagonalization representation of Hamiltonian $\hat{H}_{TC}$. In order to facilitate the discussion, we consider the resonance condition $\omega_{a}=\omega_{c}$ (i.e., $\Delta=0$) in the following analysis.  

The $M$-particle eigenvectors of the operator $\hat{t}(\lambda)$ are constructed according to algebraic BA
\begin{equation}
\left|\Phi_{J,M}\left(\{\lambda\}\right)\right\rangle =\prod_{j=1}^{M}\hat{X}_{+}(\lambda_{j})\left|\widetilde{0}\right\rangle,                    \label{A4} 
\end{equation}
where $\{\lambda\}=\{\lambda_{1},\lambda_{2}, \cdot, \lambda_{M}\}$, $\hat{X}_{+}(\lambda_{j})$ are quasi-particle creation operators, and $\left|\widetilde{0}\right\rangle =\left|0\right\rangle\otimes \left|J,-J\right\rangle $ is a pseudo-vaccum state, $\hat{a}\left|0\right\rangle=0$, $\hat{J}_{-}\left|J,-J\right\rangle =0$ ($0\leq J\leq N/2$), and $\hat{X}_{-}(\lambda)|\widetilde{0}\rangle=0$. The eigenequation of $\hat{t}(\mu)$ is           
\begin{eqnarray}
\hat{t}(\mu)\left|\Phi_{J,M}\left(\{\lambda\}\right)\right\rangle =\varTheta_{M}(\mu)\left|\Phi_{J,M}\left(\{\lambda\}\right)\right\rangle \nonumber\\ +2\sum_{j=1}^{M}\frac{f_{j}}{\mu-\lambda_{j}}\hat{X}^{+}(\text{\ensuremath{\mu}})\left|\Phi_{J,M}\left(\{\lambda\}\right)\right\rangle, \label{A5}
\end{eqnarray}
where $\mu$ is a complex number, and
\begin{eqnarray}
\varTheta_{M}(\mu)&=&-2\sum_{j=1}^{M}\frac{1}{\mu-\lambda_{j}}\left[\frac{J}{\mu}-\frac{\mu}{2}-\sum_{k=1,k\neq j}^{M}\frac{1}{\lambda_{j}-\lambda_{k}}\right]   \nonumber \\
&&+\mu^{-2}J(J+1)+(\frac{1}{2}-J)+\frac{\mu^{2}}{4} ,  \label{A6} \\
f_{j}&=&\frac{J}{\lambda_{j}}-\frac{\text{\ensuremath{\lambda_{j}}}}{2}-\sum_{k=1,k\neq j}^{M}\frac{1}{\lambda_{j}-\lambda_{k}} . \label{A7}
\end{eqnarray}
If all $f_{j}=0$, then $\left|\Phi_{J,M}\left(\{\lambda\}\right)\right\rangle$  is an eigenvector of $\hat{t}(\mu)$, and the equation
\begin{equation}
\frac{J}{\lambda_{j}}-\frac{\text{\ensuremath{\lambda_{j}}}}{2}-\sum_{k=1,k\neq j}^{M}\frac{1}{\lambda_{j}-\lambda_{k}}=0,  \label{A8}
\end{equation}
are known as the Bethe ansatz equation (BAE). There are $K=min(2J, M)+1$ sets of solutions of these BAE, and the complex-valued roots are pairwise conjugated \cite{vladimirov1986proof}. The corresponding eigenvalues of $\hat{t}(\mu)$ are $\varTheta_{M}(\mu)$.

The $\sigma$-th($\sigma\leq K$) eigenequaiton of $\hat{H}_{TC}$
\begin{equation}
\hat{H}_{TC}\left|\Phi_{J,M}\left(\{\lambda^{\sigma}\}\right)\right\rangle  =E_{J,M}^{\sigma}\left|\Phi_{J,M}\left(\{\lambda^{\sigma}\}\right)\right\rangle ,  \label{A9}
\end{equation}
where $\{\lambda^{\sigma}\}$ is the $\sigma$-th set of solution of BAE and the  corresponding eigenvalue is $E_{J,M}^{\sigma}=-\sum_{j}^{M}\lambda_{j}$.

\section{\label{Appendix B}The calculation of Eq. (\ref{7})}
In this Appendix, we present the calculation of the inner product $C\left(\Phi_{J,M}^{\sigma},\Phi(0)\right)=\left\langle \Phi_{J,M}\left(\{\lambda^{\sigma}\}\right)\right|\left.\Phi(0)\right\rangle $.
Substituting Eq. (\ref{3}) and Eq. (\ref{A4}) into $C\left(\Phi_{J,M}^{\sigma},\Phi(0)\right)$ and inserting the unit operator $\hat{I}=\sum_{n=0}^{\infty}\left|n\right\rangle \left\langle n\right|\otimes\sum_{s=-J}^{J}\left|J,s\right\rangle \left\langle J,s\right|$, then (In order to prevent the formula from escaping the page, we will replace $C\left(\Phi_{J,M}^{\sigma},\Phi(0)\right)$ with $C$ below.)
{\allowdisplaybreaks
\begin{eqnarray}
C&=&\left\langle J,-J\right|\otimes\left\langle 0\right|\prod_{j=1}^{M}\left(\hat{a}-\frac{\hat{J}_{-}}{\lambda_{j}^{\sigma}}\right)\sum_{n=0}^{\infty}\left|n\right\rangle \left\langle n\right| \nonumber\\ &&\otimes\sum_{s=-J}^{J}\left|J,s\right\rangle \left\langle J,s\right|\left|\psi_{0}\right\rangle \otimes\left|J,-J\right\rangle  \nonumber \\
&=&\left\langle J,-J\right|\otimes\left\langle 0\right|\sum_{n=0}^{\infty}C(n,\psi_{0})\prod_{j=1}^{M-1}\left(\hat{a}-\frac{\hat{J}_{-}}{\lambda_{j}^{\sigma}}\right) \nonumber \\
&& \times\sqrt{n}\left|n-1\right\rangle \left|J,-J\right\rangle \nonumber\\
&=&\left\langle J,-J\right|\otimes\left\langle 0\right|\sum_{n=0}^{\infty}C(n,\psi_{0})\sqrt{\frac{n!}{(n-M)!}} \nonumber \\
&&\times\left|n-M\right\rangle \otimes\left|J,-J\right\rangle \nonumber \\
&=&C(M,\psi_{0})\sqrt{M!}   ,\label{B1}  
\end{eqnarray}}
where $C(M,\psi_{0})=\left. \langle M \right. \left|\psi_{0}\right\rangle$ is the inner product of the arbitrary initial state $\left|\psi_{0}\right\rangle$ and the number state $\left|M\right\rangle$ of the cavity field.

\section{\label{Appendix C}Some solutions of Bethe ansatz equation and expressions of function $F(M,t)$}
Since the solutions of BAE and the expressions of the function $F(M,t)$ are too cumbersome, we only show some of them here. These solutions in table $\rm\uppercase\expandafter{\romannumeral 1}$ are obtained in Eq. (\ref{A8}) when $N = 10$.

{\allowdisplaybreaks
\begin{table}[!h]
\caption{\label{table1}
Some solutions of BAE(Eq. (\ref{A8})).}
\begin{ruledtabular}
\begin{tabular}{cccccccc}
&M=1 &M=2 &M=3&M=4 \\
\hline
$\{\lambda_{i}^{(1)}\}$&3.16 & $3.08\pm0.7i$ & -2.83 &2.68,-2.88 \\
& & & $-2.83\pm0.70i$&  $-2.79\pm1.22i$ \\
\hline
$\{\lambda_{i}^{(2)}\}$& -3.16 & $-3.08\pm0.7i$ & 2.83 &-2.68,2.88\\
&&&$2.83\pm0.70i$&$2.79\pm1.22i$\\
\hline
$\{\lambda_{i}^{(3)}\}$&  & $\pm3$ & -3.06 &$2.99\pm0.53i$ \\
&&&$-2.97\pm1.23i$&$2.86\pm1.68i$ \\
\hline
$\{\lambda_{i}^{(4)}\}$&&&3.06& $-2.99\pm0.53i$ \\
&&&$2.97\pm1.23i$&$-2.86\pm1.68i$  \\
\hline
$\{\lambda_{i}^{(5)}\}$&&&&$2.73\pm 0.68i$ \\
&&&&$-2.73\pm 0.68i$
\end{tabular}
\end{ruledtabular}
\end{table}}
Some specific expressions of the number state stored energy are shown below. Special attention should be paid to the fact that we ignored high-order small quantities in the following expressions
{\allowdisplaybreaks
\begin{subequations}
\begin{eqnarray}
F(0,t)&=&0 ,\label{C1a}\\
F(1,t)&=&1-\cos^{2}\left(\sqrt{10}t\right) ,\label{C1b}\\
F(2,t)&=&1.01-\cos(6.16t)-0.01\cos(12.33t) ,\label{C1c}\\
F(3,t)&=&1.53-0.71\cos(5.96t)-0.04\cos(11.99t) \nonumber\\
&&-0.78\cos(6.03t) ,\label{C1d}\\
F(4,t)&=&2.07-0.04\cos(11.54t)-1.40\cos(5.77t) \nonumber\\
&&-0.58\cos(5.92t)-0.05\cos(11.69t) ,\label{C1e}\\
F(5,t)&=&2.63-0.05\cos(11.46t)-0.42\cos(5.85t) \nonumber\\
&&-0.82\cos(5.52t)-0.12\cos(11.13t) \nonumber\\
&&-1.22\cos(5.61t) ,\label{C1f}\\
F(6,t)&=&3.19-0.13\cos(10.78t)-0.01\cos(16.07t) \nonumber\\ &&-0.31\cos(5.83t)-1.01\cos(5.49t) \nonumber\\
&&-0.04\cos(11.32t)-0.09\cos(10.58t)  \nonumber \\
&&-1.60\cos(5.29t) ,\label{C1g}\\
F(14,t)&=&5.48-0.15\cos(11.53t)-0.03\cos(17.48t) \nonumber\\
&&-1.53\cos(5.95t)-0.15\cos(12.19t) \nonumber\\
&&-0.06\cos(12.84t)-0.93\cos(6.25t) \nonumber\\
&&-0.37\cos(6.60t)-0.07\cos(6.97t) \nonumber\\
&&-0.02\cos(17.96t)-1.91\cos(5.77t) \nonumber \\
&&-0.25\cos(11.71t)-0.01\cos(13.57t) .\label{C1h}
\end{eqnarray}
\end{subequations}}

\section{\label{Appendix D}A proof of the optimal initial state}

In this Appendix, to find the optimal initial state, we propose an equal probability and equal expected value splitting method. The random variable $M$ taking non-negative integer values, $F(M)$ is a function of $M$, and $p(M)$ is an arbitrary probability distribution of $M$. Since we study the effect of different initial states of the optical field on the energy storage and average charging power of the TCQB at the same moment, we use $F(M)$ instead of $F(M,t)$ here. There are two restrictions, 
{\allowdisplaybreaks
\begin{eqnarray}
\sum_{M=0}^{M_{m}}p(M)&=&1,  \label{D1} \\
\sum_{M=0}^{M_{m}}Mp(M)&=&\bar{n}, \label{D2}
\end{eqnarray}}
where $M_{m}$ is the maximum value among the variables $M$ whose probability is not zero. $p_{s}(M)=[1-(\bar{n}-[\bar{n}])]\delta_{M,[\bar{n}]}+(\bar{n}-[\bar{n}])\delta_{M,[\bar{n}]+1}$ is a special probability distribution of $M$, where $[\bar{n}]$ indicates the integral part of $\bar{n}$. Obviously, $p_{s}(M)$ satisfies the previous two restrictions. By our proposed equal probability and equal expected value splitting method, we prove the following theory. 

If the average slope of $F(M)$ is getting smaller, then 
\begin{equation}
\sum_{M=0}^{M_{m}}p_{s}(M)F(M)\geq \sum_{M=0}^{M_{m}}p(M)F(M). \label{D3}
\end{equation} 
In contrast, if the average slope of $F(M)$ is getting larger, then 
\begin{equation}
\sum_{M=0}^{M_{m}}p_{s}(M)F(M)\leq \sum_{M=0}^{M_{m}}p(M)F(M),  \label{D4}
\end{equation} 
where $\sum_{M=0}^{M_{m}}p_{s}(M)F(M)=F([\bar{n}])+(\bar{n}-[\bar{n}])[F([\bar{n}]+1)-F([\bar{n}])]$. If $\bar{n}$ is an integer, then the last two inequalities reduce to Jensen's inequalities. That is, $F([\bar{n}])\geq \sum_{M=0}^{M_{m}}p(M)F(M)$ and $F([\bar{n}])\leq \sum_{M=0}^{M_{m}}p(M)F(M)$ when $F(M)$ is a concave function and a convex function, respectively. If $F(M)$ is a convex function, its average slope is getting larger. In contrast, if $F(M)$ is a concave function, its average slope gets smaller and smaller. We will explain later why we use the expression of the average slope of the function to describe the properties of the function.
  
Next, we prove the above theorem by the equal probability and equal expected value splitting method. The difference between the expected values of $F(M)$ under the two probability distributions of $p_{s}(M)$ and $p(M)$ is as follows
{\allowdisplaybreaks
\begin{eqnarray}
\varDelta F&=&\sum_{M=0}^{M_{m}}p_{s}(M)F(M)-\sum_{M=0}^{M_{m}}p(M)F(M) \nonumber \\
&=&F([\bar{n}])-\sum_{M=0}^{M_{m}}p(M)F(M)+(\bar{n}-[\bar{n}])[F([\bar{n}]+1) \nonumber \\
&&-F([\bar{n}])].   \label{D5}
\end{eqnarray}}
Substituting $\sum_{M=0}^{M_{m}}p(M)=1$ into the above equation, we obtain
{\allowdisplaybreaks
\begin{eqnarray}
\varDelta F&=&\sum_{M=0}^{M_{m}}p(M)[F([\bar{n}])-F(M)]+(\bar{n}-[\bar{n}])[F([\bar{n}]+1) \nonumber \\
&&-F([\bar{n}])] \nonumber \\
&=&\sum_{i=0}^{[\bar{n}]-1}p(i)[F([\bar{n}])-F(i)]+\sum_{j=1}^{d}p([\bar{n}]+j)[F([\bar{n}]) \nonumber \\
&&-F([\bar{n}]+j)]+(\bar{n}-[\bar{n}])[F([\bar{n}]+1)-F([\bar{n}])],  \label{D6}
\end{eqnarray}}
where $d=M_{m}-[\bar{n}]$.

We rewrite Eq. (\ref{D1}) and equation Eq. (\ref{D2}) in the following forms
\begin{eqnarray}
&&\sum_{i=0}^{[\bar{n}]-1}p(i)+p([\bar{n}])+\sum_{j=1}^{d}p([\bar{n}]+j)=1,  \label{D7} \\
&&\sum_{i=0}^{[\bar{n}]-1}p(i)i+p([\bar{n}])[\bar{n}]+\sum_{j=1}^{d}p([\bar{n}]+j)([\bar{n}]+j)=\bar{n}. \label{D8}
\end{eqnarray}
From the above two equations, we can obtain the following expression
\begin{equation}
\sum_{i=0}^{[\bar{n}]-1}p(i)([\bar{n}]-i)+(\bar{n}-[\bar{n}])=\sum_{j=1}^{d}p([\bar{n}]+j)j. \label{D9}
\end{equation}
We divide $p(i)$, $(\bar{n}-[\bar{n}])$ and $[1-(\bar{n}-[\bar{n}])]$ into $d$ parts, and their $j$-th parts are respectively as follows
{\allowdisplaybreaks
\begin{eqnarray}
p_{j}(i)&=&\frac{jp([\bar{n}]+j)}{\sum_{k=1}^{d}kp([\bar{n}]+k)}p(i), \label{D10}\\
(\bar{n}-[\bar{n}])_{j}&=&\frac{jp([\bar{n}]+j)}{\sum_{k=1}^{d}kp([\bar{n}]+k)}(\bar{n}-[\bar{n}]), \label{D11}\\
\lbrack 1-(\bar{n}-[\bar{n}])\rbrack_{j}&=&\frac{jp([\bar{n}]+j)}{\sum_{k=1}^{d}kp([\bar{n}]+k)}\Bigg[\sum_{i=0}^{[\bar{n}]}p(i)-(\bar{n}-[\bar{n}])\Bigg] \nonumber \\
&&+p(\lbrack\bar{n}\rbrack+j). \label{D12}
\end{eqnarray}}
It is easy to verify that $\sum_{j=1}^{d}p_{j}(i)=p(i)$, $\sum_{j=1}^{d}(\bar{n}-[\bar{n}])_{j}=(\bar{n}-[\bar{n}])$ and $\sum_{j=1}^{d}[1-(\bar{n}-[\bar{n}])]_{j}=[1-(\bar{n}-[\bar{n}])]$. We can rewrite Eq. (\ref{D9}) in the following form
{\allowdisplaybreaks
\begin{eqnarray}
&&\sum_{i=0}^{[\bar{n}]-1}p(i)([\bar{n}]-i)+(\bar{n}-[\bar{n}]) \nonumber \\
&&=\sum_{i=0}^{[\bar{n}]-1}\sum_{j=1}^{d}p_{j}(i)([\bar{n}]-i)+\sum_{j=1}^{d}(\bar{n}-[\bar{n}])_{j}  \nonumber \\
&&=\sum_{j=1}^{d}\Bigg[\sum_{i=0}^{[\bar{n}]-1}p_{j}(i)([\bar{n}]-i)+(\bar{n}-[\bar{n}])_{j}\Bigg]  \nonumber \\
&&=\sum_{j=1}^{d}p([\bar{n}]+j)j.  \label{D13}
\end{eqnarray}}
From Eq. (\ref{D10}) and Eq. (\ref{D11}), it is easy to verify that $\sum_{i=0}^{[\bar{n}]-1}p_{j}(i)([\bar{n}]-i)+(\bar{n}-[\bar{n}])_{j}=p([\bar{n}]+j)j$. Thus we can obtain the following expression
\begin{equation}
p([\bar{n}]+j)=\frac{1}{j}\Bigg[\sum_{i=0}^{[\bar{n}]-1}p_{j}(i)([\bar{n}]-i)+(\bar{n}-[\bar{n}])_{j}\Bigg].\label{D14}
\end{equation}

It is worth noting here that we were able to obtain this expression for $p([\bar{n}]+j)$ above because of Eq. (\ref{D10}) and Eq. (\ref{D11}). From Eq. (\ref{D10}), Eq. (\ref{D11}), and Eq. (\ref{D12}), the following two equations hold
{\allowdisplaybreaks
\begin{eqnarray}
&&\sum_{i=0}^{[\bar{n}]}p_{j}(i)+p([\bar{n}]+j)\nonumber \\
&=&[1-(\bar{n}-[\bar{n}])]_{j}+(\bar{n}-[\bar{n}])_{j}, \label{D15}\\
&&\sum_{i=0}^{[\bar{n}]}p_{j}(i)i+p([\bar{n}]+j)([\bar{n}]+j) \nonumber \\ 
&=&[\bar{n}][1-(\bar{n}-[\bar{n}])]_{j}+([\bar{n}]+1)(\bar{n}-[\bar{n}])_{j}.\label{D16}
\end{eqnarray} }
Eq. (\ref{D15}) indicates that after dividing the arbitrary probability distribution $p(M)$ and the special probability distribution $p_{s}(M)$ into $d$ groups, their $j$-th group probabilities are equal. Eq. (\ref{D16}) represents the equal expected value of $M$ under this equal probability split. It is the reason that we call this method as the equal probability and equal expected value splitting method.

Substituting Eq. (\ref{D14}) into Eq. (\ref{D6}) yields
{\allowdisplaybreaks
\begin{eqnarray}
\varDelta F&=&\sum_{i=0}^{[\bar{n}]-1}\sum_{j=1}^{d}p_{j}(i)([\bar{n}]-i)\Bigg[\frac{F([\bar{n}])-F(i)}{[\bar{n}]-i} \nonumber \\  
&&-\frac{F([\bar{n}]+j)-F([\bar{n}])}{j}\Bigg]+\sum_{j=1}^{d}(\bar{n}-[\bar{n}])_{j}\Bigg[\lbrack F([\bar{n}]+1) \nonumber \\
&&-F([\bar{n}])\rbrack-\frac{F([\bar{n}]+j)-F([\bar{n}])}{j}\Bigg] .\label{D17}
\end{eqnarray}}
Since $p_{j}(i)\geq0$, $([\bar{n}]-i)>0$, and $(\bar{n}-[\bar{n}])_{j}\geq0$, if the average slope of $F(M)$ is increasing, then $\varDelta F\leq0$, i.e., 
\begin{equation}
F([\bar{n}])+(\bar{n}-[\bar{n}])[F([\bar{n}]+1)-F([\bar{n}])]\leq \sum_{M=0}^{M_{m}}p(M)F(M).\label{D18}
\end{equation}
Conversely, if the average slope of $F(M)$ is decreasing, then $\varDelta F\geq0$, i.e., 
\begin{equation}
F([\bar{n}])+(\bar{n}-[\bar{n}])[F([\bar{n}]+1)-F([\bar{n}])]\geq \sum_{M=0}^{M_{m}}p(M)F(M).\label{D19}
\end{equation}
When $\bar{n}$ is an integer, the two inequalities we proved can be reduced to Jensen's inequalities.

In Fig. \ref{3}(c) and Fig. \ref{6}, we can find that the average slope of $F(M,t)$  at any time decreases or remains constant as $M$ increases in the time range we consider. Therefore, for the equal initial average photon number $\bar{n}$, the optimal initial state is the number state $\left| \bar{n} \right\rangle$ when the initial average photon number $\bar{n}$ is an integer. However, when the initial average photon number $\bar{n}$ is a non-integer, the optimal initial state is $\sqrt{(1-(\bar{n}-[\bar{n}]))}\left| [\bar{n}] \right\rangle+\sqrt{(\bar{n}-[\bar{n}])}\left| [\bar{n}]+1 \right\rangle$, which is the superposition of the number states $\left| [\bar{n}] \right\rangle$ and $\left| [\bar{n}]+1 \right\rangle$.

\nocite{*}
\bibliography{TCQB}

\providecommand{\noopsort}[1]{}\providecommand{\singleletter}[1]{#1}%
\begin{thebibliography}{48}%
\makeatletter
\providecommand \@ifxundefined [1]{%
 \@ifx{#1\undefined}
}%
\providecommand \@ifnum [1]{%
 \ifnum #1\expandafter \@firstoftwo
 \else \expandafter \@secondoftwo
 \fi
}%
\providecommand \@ifx [1]{%
 \ifx #1\expandafter \@firstoftwo
 \else \expandafter \@secondoftwo
 \fi
}%
\providecommand \natexlab [1]{#1}%
\providecommand \enquote  [1]{``#1''}%
\providecommand \bibnamefont  [1]{#1}%
\providecommand \bibfnamefont [1]{#1}%
\providecommand \citenamefont [1]{#1}%
\providecommand \href@noop [0]{\@secondoftwo}%
\providecommand \href [0]{\begingroup \@sanitize@url \@href}%
\providecommand \@href[1]{\@@startlink{#1}\@@href}%
\providecommand \@@href[1]{\endgroup#1\@@endlink}%
\providecommand \@sanitize@url [0]{\catcode `\\12\catcode `\$12\catcode
  `\&12\catcode `\#12\catcode `\^12\catcode `\_12\catcode `\%12\relax}%
\providecommand \@@startlink[1]{}%
\providecommand \@@endlink[0]{}%
\providecommand \url  [0]{\begingroup\@sanitize@url \@url }%
\providecommand \@url [1]{\endgroup\@href {#1}{\urlprefix }}%
\providecommand \urlprefix  [0]{URL }%
\providecommand \Eprint [0]{\href }%
\providecommand \doibase [0]{https://doi.org/}%
\providecommand \selectlanguage [0]{\@gobble}%
\providecommand \bibinfo  [0]{\@secondoftwo}%
\providecommand \bibfield  [0]{\@secondoftwo}%
\providecommand \translation [1]{[#1]}%
\providecommand \BibitemOpen [0]{}%
\providecommand \bibitemStop [0]{}%
\providecommand \bibitemNoStop [0]{.\EOS\space}%
\providecommand \EOS [0]{\spacefactor3000\relax}%
\providecommand \BibitemShut  [1]{\csname bibitem#1\endcsname}%
\let\auto@bib@innerbib\@empty
\bibitem [{\citenamefont {Alicki}\ and\ \citenamefont
  {Fannes}(2013)}]{alicki2013entanglement}%
  \BibitemOpen
  \bibfield  {author} {\bibinfo {author} {\bibfnamefont {R.}~\bibnamefont
  {Alicki}}\ and\ \bibinfo {author} {\bibfnamefont {M.}~\bibnamefont
  {Fannes}},\ }\bibfield  {title} {\bibinfo {title} {Entanglement boost for
  extractable work from ensembles of quantum batteries},\ }\href
  {https://doi.org/10.1103/PhysRevE.87.042123} {\bibfield  {journal} {\bibinfo
  {journal} {Phys. Rev. E}\ }\textbf {\bibinfo {volume} {87}},\ \bibinfo
  {pages} {042123} (\bibinfo {year} {2013})}\BibitemShut {NoStop}%
\bibitem [{\citenamefont {Hovhannisyan}\ \emph {et~al.}(2013)\citenamefont
  {Hovhannisyan}, \citenamefont {Perarnau-Llobet}, \citenamefont {Huber},\ and\
  \citenamefont {Ac\'{\i}n}}]{hovhannisyan2013entanglement}%
  \BibitemOpen
  \bibfield  {author} {\bibinfo {author} {\bibfnamefont {K.~V.}\ \bibnamefont
  {Hovhannisyan}}, \bibinfo {author} {\bibfnamefont {M.}~\bibnamefont
  {Perarnau-Llobet}}, \bibinfo {author} {\bibfnamefont {M.}~\bibnamefont
  {Huber}},\ and\ \bibinfo {author} {\bibfnamefont {A.}~\bibnamefont
  {Ac\'{\i}n}},\ }\bibfield  {title} {\bibinfo {title} {Entanglement generation
  is not necessary for optimal work extraction},\ }\href
  {https://doi.org/10.1103/PhysRevLett.111.240401} {\bibfield  {journal}
  {\bibinfo  {journal} {Phys. Rev. Lett.}\ }\textbf {\bibinfo {volume} {111}},\
  \bibinfo {pages} {240401} (\bibinfo {year} {2013})}\BibitemShut {NoStop}%
\bibitem [{\citenamefont {Campaioli}\ \emph {et~al.}(2017)\citenamefont
  {Campaioli}, \citenamefont {Pollock}, \citenamefont {Binder}, \citenamefont
  {C\'eleri}, \citenamefont {Goold}, \citenamefont {Vinjanampathy},\ and\
  \citenamefont {Modi}}]{campaioli2017enhancing}%
  \BibitemOpen
  \bibfield  {author} {\bibinfo {author} {\bibfnamefont {F.}~\bibnamefont
  {Campaioli}}, \bibinfo {author} {\bibfnamefont {F.~A.}\ \bibnamefont
  {Pollock}}, \bibinfo {author} {\bibfnamefont {F.~C.}\ \bibnamefont {Binder}},
  \bibinfo {author} {\bibfnamefont {L.}~\bibnamefont {C\'eleri}}, \bibinfo
  {author} {\bibfnamefont {J.}~\bibnamefont {Goold}}, \bibinfo {author}
  {\bibfnamefont {S.}~\bibnamefont {Vinjanampathy}},\ and\ \bibinfo {author}
  {\bibfnamefont {K.}~\bibnamefont {Modi}},\ }\bibfield  {title} {\bibinfo
  {title} {Enhancing the charging power of quantum batteries},\ }\href
  {https://doi.org/10.1103/PhysRevLett.118.150601} {\bibfield  {journal}
  {\bibinfo  {journal} {Phys. Rev. Lett.}\ }\textbf {\bibinfo {volume} {118}},\
  \bibinfo {pages} {150601} (\bibinfo {year} {2017})}\BibitemShut {NoStop}%
\bibitem [{\citenamefont {Manzano}\ \emph {et~al.}(2018)\citenamefont
  {Manzano}, \citenamefont {Plastina},\ and\ \citenamefont
  {Zambrini}}]{manzano2018optimal}%
  \BibitemOpen
  \bibfield  {author} {\bibinfo {author} {\bibfnamefont {G.}~\bibnamefont
  {Manzano}}, \bibinfo {author} {\bibfnamefont {F.}~\bibnamefont {Plastina}},\
  and\ \bibinfo {author} {\bibfnamefont {R.}~\bibnamefont {Zambrini}},\
  }\bibfield  {title} {\bibinfo {title} {Optimal work extraction and
  thermodynamics of quantum measurements and correlations},\ }\href
  {https://doi.org/10.1103/PhysRevLett.121.120602} {\bibfield  {journal}
  {\bibinfo  {journal} {Phys. Rev. Lett.}\ }\textbf {\bibinfo {volume} {121}},\
  \bibinfo {pages} {120602} (\bibinfo {year} {2018})}\BibitemShut {NoStop}%
\bibitem [{\citenamefont {Andolina}\ \emph
  {et~al.}(2019{\natexlab{a}})\citenamefont {Andolina}, \citenamefont {Keck},
  \citenamefont {Mari}, \citenamefont {Giovannetti},\ and\ \citenamefont
  {Polini}}]{andolina2019quantum}%
  \BibitemOpen
  \bibfield  {author} {\bibinfo {author} {\bibfnamefont {G.~M.}\ \bibnamefont
  {Andolina}}, \bibinfo {author} {\bibfnamefont {M.}~\bibnamefont {Keck}},
  \bibinfo {author} {\bibfnamefont {A.}~\bibnamefont {Mari}}, \bibinfo {author}
  {\bibfnamefont {V.}~\bibnamefont {Giovannetti}},\ and\ \bibinfo {author}
  {\bibfnamefont {M.}~\bibnamefont {Polini}},\ }\bibfield  {title} {\bibinfo
  {title} {Quantum versus classical many-body batteries},\ }\href
  {https://doi.org/10.1103/PhysRevB.99.205437} {\bibfield  {journal} {\bibinfo
  {journal} {Phys. Rev. B}\ }\textbf {\bibinfo {volume} {99}},\ \bibinfo
  {pages} {205437} (\bibinfo {year} {2019}{\natexlab{a}})}\BibitemShut
  {NoStop}%
\bibitem [{\citenamefont {Zhang}\ \emph {et~al.}(2019)\citenamefont {Zhang},
  \citenamefont {Yang}, \citenamefont {Fu},\ and\ \citenamefont
  {Wang}}]{zhang2019powerful}%
  \BibitemOpen
  \bibfield  {author} {\bibinfo {author} {\bibfnamefont {Y.-Y.}\ \bibnamefont
  {Zhang}}, \bibinfo {author} {\bibfnamefont {T.-R.}\ \bibnamefont {Yang}},
  \bibinfo {author} {\bibfnamefont {L.}~\bibnamefont {Fu}},\ and\ \bibinfo
  {author} {\bibfnamefont {X.}~\bibnamefont {Wang}},\ }\bibfield  {title}
  {\bibinfo {title} {Powerful harmonic charging in a quantum battery},\ }\href
  {https://doi.org/10.1103/PhysRevE.99.052106} {\bibfield  {journal} {\bibinfo
  {journal} {Phys. Rev. E}\ }\textbf {\bibinfo {volume} {99}},\ \bibinfo
  {pages} {052106} (\bibinfo {year} {2019})}\BibitemShut {NoStop}%
\bibitem [{\citenamefont {Yang}\ \emph {et~al.}(2020)\citenamefont {Yang},
  \citenamefont {Zhang}, \citenamefont {Dong}, \citenamefont {Fu},\ and\
  \citenamefont {Wang}}]{yang2020optimal}%
  \BibitemOpen
  \bibfield  {author} {\bibinfo {author} {\bibfnamefont {T.-R.}\ \bibnamefont
  {Yang}}, \bibinfo {author} {\bibfnamefont {Y.-Y.}\ \bibnamefont {Zhang}},
  \bibinfo {author} {\bibfnamefont {H.}~\bibnamefont {Dong}}, \bibinfo {author}
  {\bibfnamefont {L.}~\bibnamefont {Fu}},\ and\ \bibinfo {author}
  {\bibfnamefont {X.}~\bibnamefont {Wang}},\ }\bibfield  {title} {\bibinfo
  {title} {Optimal building block of multipartite quantum battery},\
  }\href@noop {} {\bibfield  {journal} {\bibinfo  {journal} {arXiv preprint
  arXiv:2010.09970}\ } (\bibinfo {year} {2020})}\BibitemShut {NoStop}%
\bibitem [{\citenamefont {Chen}\ \emph {et~al.}(2020)\citenamefont {Chen},
  \citenamefont {Zhan}, \citenamefont {Shao}, \citenamefont {Zhang},
  \citenamefont {Zhang},\ and\ \citenamefont {Wang}}]{chen2020charging}%
  \BibitemOpen
  \bibfield  {author} {\bibinfo {author} {\bibfnamefont {J.}~\bibnamefont
  {Chen}}, \bibinfo {author} {\bibfnamefont {L.}~\bibnamefont {Zhan}}, \bibinfo
  {author} {\bibfnamefont {L.}~\bibnamefont {Shao}}, \bibinfo {author}
  {\bibfnamefont {X.}~\bibnamefont {Zhang}}, \bibinfo {author} {\bibfnamefont
  {Y.}~\bibnamefont {Zhang}},\ and\ \bibinfo {author} {\bibfnamefont
  {X.}~\bibnamefont {Wang}},\ }\bibfield  {title} {\bibinfo {title} {Charging
  quantum batteries with a general harmonic driving field},\ }\href@noop {}
  {\bibfield  {journal} {\bibinfo  {journal} {Ann. Phys. (Berlin)}\ }\textbf
  {\bibinfo {volume} {532}},\ \bibinfo {pages} {1900487} (\bibinfo {year}
  {2020})}\BibitemShut {NoStop}%
\bibitem [{\citenamefont {Andolina}\ \emph
  {et~al.}(2019{\natexlab{b}})\citenamefont {Andolina}, \citenamefont {Keck},
  \citenamefont {Mari}, \citenamefont {Campisi}, \citenamefont {Giovannetti},\
  and\ \citenamefont {Polini}}]{PhysRevLett.122.047702}%
  \BibitemOpen
  \bibfield  {author} {\bibinfo {author} {\bibfnamefont {G.~M.}\ \bibnamefont
  {Andolina}}, \bibinfo {author} {\bibfnamefont {M.}~\bibnamefont {Keck}},
  \bibinfo {author} {\bibfnamefont {A.}~\bibnamefont {Mari}}, \bibinfo {author}
  {\bibfnamefont {M.}~\bibnamefont {Campisi}}, \bibinfo {author} {\bibfnamefont
  {V.}~\bibnamefont {Giovannetti}},\ and\ \bibinfo {author} {\bibfnamefont
  {M.}~\bibnamefont {Polini}},\ }\bibfield  {title} {\bibinfo {title}
  {Extractable work, the role of correlations, and asymptotic freedom in
  quantum batteries},\ }\href {https://doi.org/10.1103/PhysRevLett.122.047702}
  {\bibfield  {journal} {\bibinfo  {journal} {Phys. Rev. Lett.}\ }\textbf
  {\bibinfo {volume} {122}},\ \bibinfo {pages} {047702} (\bibinfo {year}
  {2019}{\natexlab{b}})}\BibitemShut {NoStop}%
\bibitem [{\citenamefont {Ferraro}\ \emph {et~al.}(2018)\citenamefont
  {Ferraro}, \citenamefont {Campisi}, \citenamefont {Andolina}, \citenamefont
  {Pellegrini},\ and\ \citenamefont {Polini}}]{ferraro2018high}%
  \BibitemOpen
  \bibfield  {author} {\bibinfo {author} {\bibfnamefont {D.}~\bibnamefont
  {Ferraro}}, \bibinfo {author} {\bibfnamefont {M.}~\bibnamefont {Campisi}},
  \bibinfo {author} {\bibfnamefont {G.~M.}\ \bibnamefont {Andolina}}, \bibinfo
  {author} {\bibfnamefont {V.}~\bibnamefont {Pellegrini}},\ and\ \bibinfo
  {author} {\bibfnamefont {M.}~\bibnamefont {Polini}},\ }\bibfield  {title}
  {\bibinfo {title} {High-power collective charging of a solid-state quantum
  battery},\ }\href {https://doi.org/10.1103/PhysRevLett.120.117702} {\bibfield
   {journal} {\bibinfo  {journal} {Phys. Rev. Lett.}\ }\textbf {\bibinfo
  {volume} {120}},\ \bibinfo {pages} {117702} (\bibinfo {year}
  {2018})}\BibitemShut {NoStop}%
\bibitem [{\citenamefont {Crescente}\ \emph {et~al.}(2020)\citenamefont
  {Crescente}, \citenamefont {Carrega}, \citenamefont {Sassetti},\ and\
  \citenamefont {Ferraro}}]{crescente2020ultrafast}%
  \BibitemOpen
  \bibfield  {author} {\bibinfo {author} {\bibfnamefont {A.}~\bibnamefont
  {Crescente}}, \bibinfo {author} {\bibfnamefont {M.}~\bibnamefont {Carrega}},
  \bibinfo {author} {\bibfnamefont {M.}~\bibnamefont {Sassetti}},\ and\
  \bibinfo {author} {\bibfnamefont {D.}~\bibnamefont {Ferraro}},\ }\bibfield
  {title} {\bibinfo {title} {Ultrafast charging in a two-photon dicke quantum
  battery},\ }\href {https://doi.org/10.1103/PhysRevB.102.245407} {\bibfield
  {journal} {\bibinfo  {journal} {Phys. Rev. B}\ }\textbf {\bibinfo {volume}
  {102}},\ \bibinfo {pages} {245407} (\bibinfo {year} {2020})}\BibitemShut
  {NoStop}%
\bibitem [{\citenamefont {Le}\ \emph {et~al.}(2018)\citenamefont {Le},
  \citenamefont {Levinsen}, \citenamefont {Modi}, \citenamefont {Parish},\ and\
  \citenamefont {Pollock}}]{le2018spin}%
  \BibitemOpen
  \bibfield  {author} {\bibinfo {author} {\bibfnamefont {T.~P.}\ \bibnamefont
  {Le}}, \bibinfo {author} {\bibfnamefont {J.}~\bibnamefont {Levinsen}},
  \bibinfo {author} {\bibfnamefont {K.}~\bibnamefont {Modi}}, \bibinfo {author}
  {\bibfnamefont {M.~M.}\ \bibnamefont {Parish}},\ and\ \bibinfo {author}
  {\bibfnamefont {F.~A.}\ \bibnamefont {Pollock}},\ }\bibfield  {title}
  {\bibinfo {title} {Spin-chain model of a many-body quantum battery},\ }\href
  {https://doi.org/10.1103/PhysRevA.97.022106} {\bibfield  {journal} {\bibinfo
  {journal} {Phys. Rev. A}\ }\textbf {\bibinfo {volume} {97}},\ \bibinfo
  {pages} {022106} (\bibinfo {year} {2018})}\BibitemShut {NoStop}%
\bibitem [{\citenamefont {Peng}\ \emph {et~al.}(2021)\citenamefont {Peng},
  \citenamefont {He}, \citenamefont {Chesi}, \citenamefont {Lin},\ and\
  \citenamefont {Guan}}]{peng2021lower}%
  \BibitemOpen
  \bibfield  {author} {\bibinfo {author} {\bibfnamefont {L.}~\bibnamefont
  {Peng}}, \bibinfo {author} {\bibfnamefont {W.-B.}\ \bibnamefont {He}},
  \bibinfo {author} {\bibfnamefont {S.}~\bibnamefont {Chesi}}, \bibinfo
  {author} {\bibfnamefont {H.-Q.}\ \bibnamefont {Lin}},\ and\ \bibinfo {author}
  {\bibfnamefont {X.-W.}\ \bibnamefont {Guan}},\ }\bibfield  {title} {\bibinfo
  {title} {Lower and upper bounds of quantum battery power in multiple central
  spin systems},\ }\href {https://doi.org/10.1103/PhysRevA.103.052220}
  {\bibfield  {journal} {\bibinfo  {journal} {Phys. Rev. A}\ }\textbf {\bibinfo
  {volume} {103}},\ \bibinfo {pages} {052220} (\bibinfo {year}
  {2021})}\BibitemShut {NoStop}%
\bibitem [{\citenamefont {Caravelli}\ \emph {et~al.}(2020)\citenamefont
  {Caravelli}, \citenamefont {Coulter-DeWit}, \citenamefont
  {Garc{\'\i}a-Pintos},\ and\ \citenamefont {Hamma}}]{caravelli2020random}%
  \BibitemOpen
  \bibfield  {author} {\bibinfo {author} {\bibfnamefont {F.}~\bibnamefont
  {Caravelli}}, \bibinfo {author} {\bibfnamefont {G.}~\bibnamefont
  {Coulter-DeWit}}, \bibinfo {author} {\bibfnamefont {L.~P.}\ \bibnamefont
  {Garc{\'\i}a-Pintos}},\ and\ \bibinfo {author} {\bibfnamefont
  {A.}~\bibnamefont {Hamma}},\ }\bibfield  {title} {\bibinfo {title} {Random
  quantum batteries},\ }\href
  {https://doi.org/10.1103/PhysRevResearch.2.023095} {\bibfield  {journal}
  {\bibinfo  {journal} {Phys. Rev. Research}\ }\textbf {\bibinfo {volume}
  {2}},\ \bibinfo {pages} {023095} (\bibinfo {year} {2020})}\BibitemShut
  {NoStop}%
\bibitem [{\citenamefont {Rossini}\ \emph {et~al.}(2020)\citenamefont
  {Rossini}, \citenamefont {Andolina}, \citenamefont {Rosa}, \citenamefont
  {Carrega},\ and\ \citenamefont {Polini}}]{rossini2020quantum}%
  \BibitemOpen
  \bibfield  {author} {\bibinfo {author} {\bibfnamefont {D.}~\bibnamefont
  {Rossini}}, \bibinfo {author} {\bibfnamefont {G.~M.}\ \bibnamefont
  {Andolina}}, \bibinfo {author} {\bibfnamefont {D.}~\bibnamefont {Rosa}},
  \bibinfo {author} {\bibfnamefont {M.}~\bibnamefont {Carrega}},\ and\ \bibinfo
  {author} {\bibfnamefont {M.}~\bibnamefont {Polini}},\ }\bibfield  {title}
  {\bibinfo {title} {Quantum advantage in the charging process of
  sachdev-ye-kitaev batteries},\ }\href
  {https://doi.org/10.1103/PhysRevLett.125.236402} {\bibfield  {journal}
  {\bibinfo  {journal} {Phys. Rev. Lett.}\ }\textbf {\bibinfo {volume} {125}},\
  \bibinfo {pages} {236402} (\bibinfo {year} {2020})}\BibitemShut {NoStop}%
\bibitem [{\citenamefont {Patil}\ \emph {et~al.}(2020)\citenamefont {Patil},
  \citenamefont {Kos}, \citenamefont {Ravnik},\ and\ \citenamefont
  {Dunkel}}]{patil2020discharging}%
  \BibitemOpen
  \bibfield  {author} {\bibinfo {author} {\bibfnamefont {V.~P.}\ \bibnamefont
  {Patil}}, \bibinfo {author} {\bibfnamefont {{\v{Z}}.}~\bibnamefont {Kos}},
  \bibinfo {author} {\bibfnamefont {M.}~\bibnamefont {Ravnik}},\ and\ \bibinfo
  {author} {\bibfnamefont {J.}~\bibnamefont {Dunkel}},\ }\bibfield  {title}
  {\bibinfo {title} {Discharging dynamics of topological batteries},\ }\href
  {https://doi.org/10.1103/PhysRevResearch.2.043196} {\bibfield  {journal}
  {\bibinfo  {journal} {Phys. Rev. Research}\ }\textbf {\bibinfo {volume}
  {2}},\ \bibinfo {pages} {043196} (\bibinfo {year} {2020})}\BibitemShut
  {NoStop}%
\bibitem [{\citenamefont {Barra}(2019)}]{barra2019dissipative}%
  \BibitemOpen
  \bibfield  {author} {\bibinfo {author} {\bibfnamefont {F.}~\bibnamefont
  {Barra}},\ }\bibfield  {title} {\bibinfo {title} {Dissipative charging of a
  quantum battery},\ }\href {https://doi.org/10.1103/PhysRevLett.122.210601}
  {\bibfield  {journal} {\bibinfo  {journal} {Phys. Rev. Lett.}\ }\textbf
  {\bibinfo {volume} {122}},\ \bibinfo {pages} {210601} (\bibinfo {year}
  {2019})}\BibitemShut {NoStop}%
\bibitem [{\citenamefont {Quach}\ and\ \citenamefont
  {Munro}(2020)}]{quach2020using}%
  \BibitemOpen
  \bibfield  {author} {\bibinfo {author} {\bibfnamefont {J.~Q.}\ \bibnamefont
  {Quach}}\ and\ \bibinfo {author} {\bibfnamefont {W.~J.}\ \bibnamefont
  {Munro}},\ }\bibfield  {title} {\bibinfo {title} {Using dark states to charge
  and stabilize open quantum batteries},\ }\href
  {https://doi.org/10.1103/PhysRevApplied.14.024092} {\bibfield  {journal}
  {\bibinfo  {journal} {Phys. Rev. Applied}\ }\textbf {\bibinfo {volume}
  {14}},\ \bibinfo {pages} {024092} (\bibinfo {year} {2020})}\BibitemShut
  {NoStop}%
\bibitem [{\citenamefont {Pirmoradian}\ and\ \citenamefont
  {M\o{}lmer}(2019)}]{pirmoradian2019aging}%
  \BibitemOpen
  \bibfield  {author} {\bibinfo {author} {\bibfnamefont {F.}~\bibnamefont
  {Pirmoradian}}\ and\ \bibinfo {author} {\bibfnamefont {K.}~\bibnamefont
  {M\o{}lmer}},\ }\bibfield  {title} {\bibinfo {title} {Aging of a quantum
  battery},\ }\href {https://doi.org/10.1103/PhysRevA.100.043833} {\bibfield
  {journal} {\bibinfo  {journal} {Phys. Rev. A}\ }\textbf {\bibinfo {volume}
  {100}},\ \bibinfo {pages} {043833} (\bibinfo {year} {2019})}\BibitemShut
  {NoStop}%
\bibitem [{\citenamefont {Bai}\ and\ \citenamefont
  {An}(2020)}]{bai2020floquet}%
  \BibitemOpen
  \bibfield  {author} {\bibinfo {author} {\bibfnamefont {S.-Y.}\ \bibnamefont
  {Bai}}\ and\ \bibinfo {author} {\bibfnamefont {J.-H.}\ \bibnamefont {An}},\
  }\bibfield  {title} {\bibinfo {title} {Floquet engineering to reactivate a
  dissipative quantum battery},\ }\href
  {https://doi.org/10.1103/PhysRevA.102.060201} {\bibfield  {journal} {\bibinfo
   {journal} {Phys. Rev. A}\ }\textbf {\bibinfo {volume} {102}},\ \bibinfo
  {pages} {060201(R)} (\bibinfo {year} {2020})}\BibitemShut {NoStop}%
\bibitem [{\citenamefont {Bethe}(1931)}]{BetheH}%
  \BibitemOpen
  \bibfield  {author} {\bibinfo {author} {\bibfnamefont {H.}~\bibnamefont
  {Bethe}},\ }\bibfield  {title} {\bibinfo {title} {Zur theorie der metalle},\
  }\href@noop {} {\bibfield  {journal} {\bibinfo  {journal} {Z. Physik}\
  }\textbf {\bibinfo {volume} {71}},\ \bibinfo {pages} {205} (\bibinfo {year}
  {1931})}\BibitemShut {NoStop}%
\bibitem [{\citenamefont {Lieb}\ and\ \citenamefont
  {Liniger}(1963)}]{lieb1963exact}%
  \BibitemOpen
  \bibfield  {author} {\bibinfo {author} {\bibfnamefont {E.~H.}\ \bibnamefont
  {Lieb}}\ and\ \bibinfo {author} {\bibfnamefont {W.}~\bibnamefont {Liniger}},\
  }\bibfield  {title} {\bibinfo {title} {Exact analysis of an interacting bose
  gas. i. the general solution and the ground state},\ }\href
  {https://doi.org/10.1103/PhysRev.130.1605} {\bibfield  {journal} {\bibinfo
  {journal} {Phys. Rev.}\ }\textbf {\bibinfo {volume} {130}},\ \bibinfo {pages}
  {1605} (\bibinfo {year} {1963})}\BibitemShut {NoStop}%
\bibitem [{\citenamefont {Lieb}(1963)}]{lieb1963exact1}%
  \BibitemOpen
  \bibfield  {author} {\bibinfo {author} {\bibfnamefont {E.~H.}\ \bibnamefont
  {Lieb}},\ }\bibfield  {title} {\bibinfo {title} {Exact analysis of an
  interacting bose gas. ii. the excitation spectrum},\ }\href
  {https://doi.org/10.1103/PhysRev.130.1616} {\bibfield  {journal} {\bibinfo
  {journal} {Phys. Rev.}\ }\textbf {\bibinfo {volume} {130}},\ \bibinfo {pages}
  {1616} (\bibinfo {year} {1963})}\BibitemShut {NoStop}%
\bibitem [{\citenamefont {Lieb}\ and\ \citenamefont
  {Wu}(1968)}]{lieb1968absence}%
  \BibitemOpen
  \bibfield  {author} {\bibinfo {author} {\bibfnamefont {E.~H.}\ \bibnamefont
  {Lieb}}\ and\ \bibinfo {author} {\bibfnamefont {F.~Y.}\ \bibnamefont {Wu}},\
  }\bibfield  {title} {\bibinfo {title} {Absence of mott transition in an exact
  solution of the short-range, one-band model in one dimension},\ }\href
  {https://doi.org/10.1103/PhysRevLett.20.1445} {\bibfield  {journal} {\bibinfo
   {journal} {Phys. Rev. Lett.}\ }\textbf {\bibinfo {volume} {20}},\ \bibinfo
  {pages} {1445} (\bibinfo {year} {1968})}\BibitemShut {NoStop}%
\bibitem [{\citenamefont {Griffiths}(1964)}]{griffiths1964magnetization}%
  \BibitemOpen
  \bibfield  {author} {\bibinfo {author} {\bibfnamefont {R.~B.}\ \bibnamefont
  {Griffiths}},\ }\bibfield  {title} {\bibinfo {title} {Magnetization curve at
  zero temperature for the antiferromagnetic heisenberg linear chain},\ }\href
  {https://doi.org/10.1103/PhysRev.133.A768} {\bibfield  {journal} {\bibinfo
  {journal} {Phys. Rev.}\ }\textbf {\bibinfo {volume} {133}},\ \bibinfo {pages}
  {A768} (\bibinfo {year} {1964})}\BibitemShut {NoStop}%
\bibitem [{\citenamefont {Bolech}\ and\ \citenamefont
  {Andrei}(2002)}]{CJBolech}%
  \BibitemOpen
  \bibfield  {author} {\bibinfo {author} {\bibfnamefont {C.~J.}\ \bibnamefont
  {Bolech}}\ and\ \bibinfo {author} {\bibfnamefont {N.}~\bibnamefont
  {Andrei}},\ }\bibfield  {title} {\bibinfo {title} {Solution of the
  two-channel anderson impurity model: Implications for the heavy fermion
  ${\mathrm{ube}}_{13}$},\ }\href
  {https://doi.org/10.1103/PhysRevLett.88.237206} {\bibfield  {journal}
  {\bibinfo  {journal} {Phys. Rev. Lett.}\ }\textbf {\bibinfo {volume} {88}},\
  \bibinfo {pages} {237206} (\bibinfo {year} {2002})}\BibitemShut {NoStop}%
\bibitem [{\citenamefont {Andrei}\ and\ \citenamefont
  {Destri}(1984)}]{NAndrei}%
  \BibitemOpen
  \bibfield  {author} {\bibinfo {author} {\bibfnamefont {N.}~\bibnamefont
  {Andrei}}\ and\ \bibinfo {author} {\bibfnamefont {C.}~\bibnamefont
  {Destri}},\ }\bibfield  {title} {\bibinfo {title} {Solution of the
  multichannel kondo problem},\ }\href
  {https://doi.org/10.1103/PhysRevLett.52.364} {\bibfield  {journal} {\bibinfo
  {journal} {Phys. Rev. Lett.}\ }\textbf {\bibinfo {volume} {52}},\ \bibinfo
  {pages} {364} (\bibinfo {year} {1984})}\BibitemShut {NoStop}%
\bibitem [{\citenamefont {Yang}(1967)}]{yang1967some}%
  \BibitemOpen
  \bibfield  {author} {\bibinfo {author} {\bibfnamefont {C.~N.}\ \bibnamefont
  {Yang}},\ }\bibfield  {title} {\bibinfo {title} {Some exact results for the
  many-body problem in one dimension with repulsive delta-function
  interaction},\ }\href {https://doi.org/10.1103/PhysRevLett.19.1312}
  {\bibfield  {journal} {\bibinfo  {journal} {Phys. Rev. Lett.}\ }\textbf
  {\bibinfo {volume} {19}},\ \bibinfo {pages} {1312} (\bibinfo {year}
  {1967})}\BibitemShut {NoStop}%
\bibitem [{\citenamefont {Baxter}(1972)}]{baxter1972partition}%
  \BibitemOpen
  \bibfield  {author} {\bibinfo {author} {\bibfnamefont {R.~J.}\ \bibnamefont
  {Baxter}},\ }\bibfield  {title} {\bibinfo {title} {Partition function of the
  eight-vertex lattice model},\ }\href
  {https://doi.org/https://doi.org/10.1016/0003-4916(72)90335-1} {\bibfield
  {journal} {\bibinfo  {journal} {Ann. Phys. (N. Y.)}\ }\textbf {\bibinfo
  {volume} {70}},\ \bibinfo {pages} {193} (\bibinfo {year} {1972})}\BibitemShut
  {NoStop}%
\bibitem [{\citenamefont {Takhtadzhan}\ and\ \citenamefont
  {Faddeev}(1979)}]{takhtadzhan1979quantum}%
  \BibitemOpen
  \bibfield  {author} {\bibinfo {author} {\bibfnamefont {L.}~\bibnamefont
  {Takhtadzhan}}\ and\ \bibinfo {author} {\bibfnamefont {L.~D.}\ \bibnamefont
  {Faddeev}},\ }\bibfield  {title} {\bibinfo {title} {The quantum method of the
  inverse problem and the heisenberg xyz model},\ }\href@noop {} {\bibfield
  {journal} {\bibinfo  {journal} {Russian Math. Surveys}\ }\textbf {\bibinfo
  {volume} {34}},\ \bibinfo {pages} {11} (\bibinfo {year} {1979})}\BibitemShut
  {NoStop}%
\bibitem [{\citenamefont {Korepin}\ \emph {et~al.}(1997)\citenamefont
  {Korepin}, \citenamefont {Bogoliubov},\ and\ \citenamefont
  {Izergin}}]{korepin1997quantum}%
  \BibitemOpen
  \bibfield  {author} {\bibinfo {author} {\bibfnamefont {V.~E.}\ \bibnamefont
  {Korepin}}, \bibinfo {author} {\bibfnamefont {N.~M.}\ \bibnamefont
  {Bogoliubov}},\ and\ \bibinfo {author} {\bibfnamefont {A.~G.}\ \bibnamefont
  {Izergin}},\ }\href@noop {} {\emph {\bibinfo {title} {Quantum inverse
  scattering method and correlation functions}}},\ Vol.~\bibinfo {volume} {3}\
  (\bibinfo  {publisher} {Cambridge university press},\ \bibinfo {year}
  {1997})\BibitemShut {NoStop}%
\bibitem [{\citenamefont {Bogoliubov}\ \emph {et~al.}(1996)\citenamefont
  {Bogoliubov}, \citenamefont {Bullough},\ and\ \citenamefont
  {Timonen}}]{bogoliubov1996exact}%
  \BibitemOpen
  \bibfield  {author} {\bibinfo {author} {\bibfnamefont {N.~M.}\ \bibnamefont
  {Bogoliubov}}, \bibinfo {author} {\bibfnamefont {R.~K.}\ \bibnamefont
  {Bullough}},\ and\ \bibinfo {author} {\bibfnamefont {J.}~\bibnamefont
  {Timonen}},\ }\bibfield  {title} {\bibinfo {title} {Exact solution of
  generalized tavis-cummings models in quantum optics},\ }\href@noop {}
  {\bibfield  {journal} {\bibinfo  {journal} {J. Phys. A: Math. Gen.}\ }\textbf
  {\bibinfo {volume} {29}},\ \bibinfo {pages} {6305} (\bibinfo {year}
  {1996})}\BibitemShut {NoStop}%
\bibitem [{\citenamefont {Bogoliubov}\ and\ \citenamefont
  {Kulish}(2013)}]{bogoliubov2013exactly}%
  \BibitemOpen
  \bibfield  {author} {\bibinfo {author} {\bibfnamefont {N.}~\bibnamefont
  {Bogoliubov}}\ and\ \bibinfo {author} {\bibfnamefont {P.}~\bibnamefont
  {Kulish}},\ }\bibfield  {title} {\bibinfo {title} {Exactly solvable models of
  quantum nonlinear optics},\ }\href@noop {} {\bibfield  {journal} {\bibinfo
  {journal} {J. Math. Sci.}\ }\textbf {\bibinfo {volume} {192}} (\bibinfo
  {year} {2013})}\BibitemShut {NoStop}%
\bibitem [{\citenamefont {Bogoliubov}\ \emph {et~al.}(2017)\citenamefont
  {Bogoliubov}, \citenamefont {Ermakov},\ and\ \citenamefont
  {Rybin}}]{bogoliubov2017time}%
  \BibitemOpen
  \bibfield  {author} {\bibinfo {author} {\bibfnamefont {N.}~\bibnamefont
  {Bogoliubov}}, \bibinfo {author} {\bibfnamefont {I.}~\bibnamefont
  {Ermakov}},\ and\ \bibinfo {author} {\bibfnamefont {A.}~\bibnamefont
  {Rybin}},\ }\bibfield  {title} {\bibinfo {title} {Time evolution of the
  atomic inversion for the generalized tavis--cummings model—qim approach},\
  }\href@noop {} {\bibfield  {journal} {\bibinfo  {journal} {J. Phys. A: Math.
  Theor.}\ }\textbf {\bibinfo {volume} {50}},\ \bibinfo {pages} {464003}
  (\bibinfo {year} {2017})}\BibitemShut {NoStop}%
\bibitem [{\citenamefont {Bates}\ \emph {et~al.}()\citenamefont {Bates},
  \citenamefont {Hauenstein}, \citenamefont {Sommese},\ and\ \citenamefont
  {Wampler}}]{BHSW06}%
  \BibitemOpen
  \bibfield  {author} {\bibinfo {author} {\bibfnamefont {D.~J.}\ \bibnamefont
  {Bates}}, \bibinfo {author} {\bibfnamefont {J.~D.}\ \bibnamefont
  {Hauenstein}}, \bibinfo {author} {\bibfnamefont {A.~J.}\ \bibnamefont
  {Sommese}},\ and\ \bibinfo {author} {\bibfnamefont {C.~W.}\ \bibnamefont
  {Wampler}},\ }\href@noop {} {\bibinfo {title} {Bertini: Software for
  numerical algebraic geometry}},\ \bibinfo {howpublished} {Available at
  bertini.nd.edu with permanent doi: dx.doi.org/10.7274/R0H41PB5}\BibitemShut
  {NoStop}%
\bibitem [{\citenamefont {Bates}\ \emph {et~al.}(2016)\citenamefont {Bates},
  \citenamefont {Newell},\ and\ \citenamefont {Niemerg}}]{bates2016bertinilab}%
  \BibitemOpen
  \bibfield  {author} {\bibinfo {author} {\bibfnamefont {D.~J.}\ \bibnamefont
  {Bates}}, \bibinfo {author} {\bibfnamefont {A.~J.}\ \bibnamefont {Newell}},\
  and\ \bibinfo {author} {\bibfnamefont {M.}~\bibnamefont {Niemerg}},\
  }\bibfield  {title} {\bibinfo {title} {Bertinilab: A matlab interface for
  solving systems of polynomial equations},\ }\href@noop {} {\bibfield
  {journal} {\bibinfo  {journal} {Numer. Algor.}\ }\textbf {\bibinfo {volume}
  {71}},\ \bibinfo {pages} {229} (\bibinfo {year} {2016})}\BibitemShut
  {NoStop}%
\bibitem [{\citenamefont {Hao}\ \emph {et~al.}(2013)\citenamefont {Hao},
  \citenamefont {Nepomechie},\ and\ \citenamefont
  {Sommese}}]{hao2013completeness}%
  \BibitemOpen
  \bibfield  {author} {\bibinfo {author} {\bibfnamefont {W.}~\bibnamefont
  {Hao}}, \bibinfo {author} {\bibfnamefont {R.~I.}\ \bibnamefont
  {Nepomechie}},\ and\ \bibinfo {author} {\bibfnamefont {A.~J.}\ \bibnamefont
  {Sommese}},\ }\bibfield  {title} {\bibinfo {title} {Completeness of solutions
  of bethe's equations},\ }\href {https://doi.org/10.1103/PhysRevE.88.052113}
  {\bibfield  {journal} {\bibinfo  {journal} {Phys. Rev. E}\ }\textbf {\bibinfo
  {volume} {88}},\ \bibinfo {pages} {052113} (\bibinfo {year}
  {2013})}\BibitemShut {NoStop}%
\bibitem [{\citenamefont {Morgan}(2009)}]{morgan2009solving}%
  \BibitemOpen
  \bibfield  {author} {\bibinfo {author} {\bibfnamefont {A.}~\bibnamefont
  {Morgan}},\ }\href@noop {} {\emph {\bibinfo {title} {Solving polynomial
  systems using continuation for engineering and scientific problems}}}\
  (\bibinfo  {publisher} {SIAM},\ \bibinfo {year} {2009})\BibitemShut {NoStop}%
\bibitem [{\citenamefont {Tavis}\ and\ \citenamefont
  {Cummings}(1968)}]{tavis1968exact}%
  \BibitemOpen
  \bibfield  {author} {\bibinfo {author} {\bibfnamefont {M.}~\bibnamefont
  {Tavis}}\ and\ \bibinfo {author} {\bibfnamefont {F.~W.}\ \bibnamefont
  {Cummings}},\ }\bibfield  {title} {\bibinfo {title} {Exact solution for an
  $n$-molecule---radiation-field hamiltonian},\ }\href
  {https://doi.org/10.1103/PhysRev.170.379} {\bibfield  {journal} {\bibinfo
  {journal} {Phys. Rev.}\ }\textbf {\bibinfo {volume} {170}},\ \bibinfo {pages}
  {379} (\bibinfo {year} {1968})}\BibitemShut {NoStop}%
\bibitem [{\citenamefont {Li}\ \emph {et~al.}(2021)\citenamefont {Li},
  \citenamefont {Braverman}, \citenamefont {Colombo}, \citenamefont {Shu},
  \citenamefont {Kawasaki}, \citenamefont {Adiyatullin}, \citenamefont
  {Pedrozo}, \citenamefont {Mendez},\ and\ \citenamefont
  {Vuleti{\'c}}}]{li2021collective}%
  \BibitemOpen
  \bibfield  {author} {\bibinfo {author} {\bibfnamefont {Z.}~\bibnamefont
  {Li}}, \bibinfo {author} {\bibfnamefont {B.}~\bibnamefont {Braverman}},
  \bibinfo {author} {\bibfnamefont {S.}~\bibnamefont {Colombo}}, \bibinfo
  {author} {\bibfnamefont {C.}~\bibnamefont {Shu}}, \bibinfo {author}
  {\bibfnamefont {A.}~\bibnamefont {Kawasaki}}, \bibinfo {author}
  {\bibfnamefont {A.}~\bibnamefont {Adiyatullin}}, \bibinfo {author}
  {\bibfnamefont {E.}~\bibnamefont {Pedrozo}}, \bibinfo {author} {\bibfnamefont
  {E.}~\bibnamefont {Mendez}},\ and\ \bibinfo {author} {\bibfnamefont
  {V.}~\bibnamefont {Vuleti{\'c}}},\ }\bibfield  {title} {\bibinfo {title}
  {Collective spin-light and light-mediated spin-spin interactions in an
  optical cavity},\ }\href@noop {} {\bibfield  {journal} {\bibinfo  {journal}
  {arXiv preprint arXiv:2106.13234}\ } (\bibinfo {year} {2021})}\BibitemShut
  {NoStop}%
\bibitem [{\citenamefont {Gegg}\ \emph {et~al.}(2018)\citenamefont {Gegg},
  \citenamefont {Carmele}, \citenamefont {Knorr},\ and\ \citenamefont
  {Richter}}]{gegg2018superradiant}%
  \BibitemOpen
  \bibfield  {author} {\bibinfo {author} {\bibfnamefont {M.}~\bibnamefont
  {Gegg}}, \bibinfo {author} {\bibfnamefont {A.}~\bibnamefont {Carmele}},
  \bibinfo {author} {\bibfnamefont {A.}~\bibnamefont {Knorr}},\ and\ \bibinfo
  {author} {\bibfnamefont {M.}~\bibnamefont {Richter}},\ }\bibfield  {title}
  {\bibinfo {title} {Superradiant to subradiant phase transition in the open
  system dicke model: Dark state cascades},\ }\href@noop {} {\bibfield
  {journal} {\bibinfo  {journal} {New Journal of Physics}\ }\textbf {\bibinfo
  {volume} {20}},\ \bibinfo {pages} {013006} (\bibinfo {year}
  {2018})}\BibitemShut {NoStop}%
\bibitem [{\citenamefont {Kirton}\ and\ \citenamefont
  {Keeling}(2017)}]{PhysRevLett.118.123602}%
  \BibitemOpen
  \bibfield  {author} {\bibinfo {author} {\bibfnamefont {P.}~\bibnamefont
  {Kirton}}\ and\ \bibinfo {author} {\bibfnamefont {J.}~\bibnamefont
  {Keeling}},\ }\bibfield  {title} {\bibinfo {title} {Suppressing and restoring
  the dicke superradiance transition by dephasing and decay},\ }\href
  {https://doi.org/10.1103/PhysRevLett.118.123602} {\bibfield  {journal}
  {\bibinfo  {journal} {Phys. Rev. Lett.}\ }\textbf {\bibinfo {volume} {118}},\
  \bibinfo {pages} {123602} (\bibinfo {year} {2017})}\BibitemShut {NoStop}%
\bibitem [{\citenamefont {Dicke}(1954)}]{dicke1954coherence}%
  \BibitemOpen
  \bibfield  {author} {\bibinfo {author} {\bibfnamefont {R.~H.}\ \bibnamefont
  {Dicke}},\ }\bibfield  {title} {\bibinfo {title} {Coherence in spontaneous
  radiation processes},\ }\href {https://doi.org/10.1103/PhysRev.93.99}
  {\bibfield  {journal} {\bibinfo  {journal} {Phys. Rev.}\ }\textbf {\bibinfo
  {volume} {93}},\ \bibinfo {pages} {99} (\bibinfo {year} {1954})}\BibitemShut
  {NoStop}%
\bibitem [{\citenamefont {Hepp}\ and\ \citenamefont
  {Lieb}(1973)}]{hepp1973superradiant}%
  \BibitemOpen
  \bibfield  {author} {\bibinfo {author} {\bibfnamefont {K.}~\bibnamefont
  {Hepp}}\ and\ \bibinfo {author} {\bibfnamefont {E.~H.}\ \bibnamefont
  {Lieb}},\ }\bibfield  {title} {\bibinfo {title} {On the superradiant phase
  transition for molecules in a quantized radiation field: the dicke maser
  model},\ }\href@noop {} {\bibfield  {journal} {\bibinfo  {journal} {Ann.
  Phys. (N. Y.)}\ }\textbf {\bibinfo {volume} {76}},\ \bibinfo {pages} {360}
  (\bibinfo {year} {1973})}\BibitemShut {NoStop}%
\bibitem [{\citenamefont {Johansson}\ \emph {et~al.}(2012)\citenamefont
  {Johansson}, \citenamefont {Nation},\ and\ \citenamefont
  {Nori}}]{johansson2012qutip}%
  \BibitemOpen
  \bibfield  {author} {\bibinfo {author} {\bibfnamefont {J.~R.}\ \bibnamefont
  {Johansson}}, \bibinfo {author} {\bibfnamefont {P.~D.}\ \bibnamefont
  {Nation}},\ and\ \bibinfo {author} {\bibfnamefont {F.}~\bibnamefont {Nori}},\
  }\bibfield  {title} {\bibinfo {title} {Qutip: An open-source python framework
  for the dynamics of open quantum systems},\ }\href@noop {} {\bibfield
  {journal} {\bibinfo  {journal} {Comput. Phys. Commun.}\ }\textbf {\bibinfo
  {volume} {183}},\ \bibinfo {pages} {1760} (\bibinfo {year}
  {2012})}\BibitemShut {NoStop}%
\bibitem [{\citenamefont {Shammah}\ \emph {et~al.}(2018)\citenamefont
  {Shammah}, \citenamefont {Ahmed}, \citenamefont {Lambert}, \citenamefont
  {De~Liberato},\ and\ \citenamefont {Nori}}]{PhysRevA.98.063815}%
  \BibitemOpen
  \bibfield  {author} {\bibinfo {author} {\bibfnamefont {N.}~\bibnamefont
  {Shammah}}, \bibinfo {author} {\bibfnamefont {S.}~\bibnamefont {Ahmed}},
  \bibinfo {author} {\bibfnamefont {N.}~\bibnamefont {Lambert}}, \bibinfo
  {author} {\bibfnamefont {S.}~\bibnamefont {De~Liberato}},\ and\ \bibinfo
  {author} {\bibfnamefont {F.}~\bibnamefont {Nori}},\ }\bibfield  {title}
  {\bibinfo {title} {Open quantum systems with local and collective incoherent
  processes: Efficient numerical simulations using permutational invariance},\
  }\href {https://doi.org/10.1103/PhysRevA.98.063815} {\bibfield  {journal}
  {\bibinfo  {journal} {Phys. Rev. A}\ }\textbf {\bibinfo {volume} {98}},\
  \bibinfo {pages} {063815} (\bibinfo {year} {2018})}\BibitemShut {NoStop}%
\bibitem [{\citenamefont {Carmichael}(1999)}]{carmichael1999statistical}%
  \BibitemOpen
  \bibfield  {author} {\bibinfo {author} {\bibfnamefont {H.~J.}\ \bibnamefont
  {Carmichael}},\ }\href@noop {} {\emph {\bibinfo {title} {Statistical methods
  in quantum optics 1: master equations and Fokker-Planck equations}}},\
  Vol.~\bibinfo {volume} {1}\ (\bibinfo  {publisher} {Springer Science \&
  Business Media},\ \bibinfo {year} {1999})\BibitemShut {NoStop}%
\bibitem [{\citenamefont {Vladimirov}(1986)}]{vladimirov1986proof}%
  \BibitemOpen
  \bibfield  {author} {\bibinfo {author} {\bibfnamefont {A.~A.}\ \bibnamefont
  {Vladimirov}},\ }\bibfield  {title} {\bibinfo {title} {Proof of the
  invariance of the bethe-ansatz solutions under complex conjugation},\
  }\href@noop {} {\bibfield  {journal} {\bibinfo  {journal} {Theor. Math.
  Phys.}\ }\textbf {\bibinfo {volume} {66}},\ \bibinfo {pages} {102} (\bibinfo
  {year} {1986})}\BibitemShut {NoStop}%
\end{thebibliography}%

\end{document}